\documentclass[twocolumn,showpacs,aps,prd,amsmath,amssymb,nofootinbib,nobibnotes,floatfix]{revtex4-1}
\usepackage{hyperref}
\usepackage{amsfonts}
\usepackage{amsmath}
\usepackage{mathrsfs}
\usepackage{epsfig}
\usepackage{graphicx}
\usepackage{url}
\usepackage{hyperref}
\usepackage{float}
\usepackage{color}
\usepackage[utf8]{inputenc} 
\usepackage{graphicx}
\usepackage{epsf}
\usepackage{comment}
\usepackage{slashed}
\usepackage{footmisc}
\usepackage{setspace}
\usepackage[dvipsnames, usenames]{xcolor}
\definecolor{linkcolor}{rgb}{0.0,0.3,0.5}

\newcommand{\de}{{\Delta{e}}}
\newcommand{\efin}{{e_{\rm fin}}}

\newcommand*{\mycdot}{\kern-.2em\cdot\kern-.2em}

\usepackage{hyperref}
\hypersetup{
    pdfnewwindow=true,      
    colorlinks=true,       
    linkcolor=linkcolor,          
    citecolor=linkcolor,        
    filecolor=blue,      
    urlcolor=blue           
}

\begin{document}

\title{The Effect of Distant Encounters on Black Hole Binaries in Globular Clusters:\\
Systematic Increase of In-cluster Mergers in the LISA band}

\author{Johan Samsing$^{1}$, Adrian S. Hamers$^{2}$, Jacob G. Tyles$^{1}$}
\affiliation{\vspace{2mm}
$^1$Department of Astrophysical Sciences, Princeton University, Peyton Hall, 4 Ivy Lane, Princeton, NJ 08544, USA.\\
$^2$Institute for Advanced Study, School of Natural Sciences, Einstein Drive, Princeton, NJ 08540, USA.
}

\begin{abstract}

We present the first systematic study on how distant weak interactions impact the dynamical evolution of merging
binary black holes (BBHs) in dense stellar clusters. Recent studies indicate that dense clusters
are likely to significantly contribute to the rate of merging BBHs observable through gravitational waves (GWs), and that many of these mergers will appear with notable eccentricities measurable in the LISA and LIGO sensitivity bands. This is highly interesting, as eccentricity can be used to distinguish between different astrophysical merger channels. However, all of these recent studies are based on various Monte Carlo (MC) techniques that only include strong interactions for the dynamical evolution of BBHs, whereas any binary generally undergoes orders-of-magnitude more weak interactions than strong. It is well known that weak interactions primarily lead to a change in the binary's eccentricity, which for BBHs implies that weak interactions can change their
GW inspiral time and thereby their merger probability.
With this motivation, we perform MC simulations of BBHs evolving in dense clusters under the influence of both
weak and strong interactions. We find that including weak interactions leads to a notable increase in the number of
BBHs that merge inside their cluster, which correspondingly leads to a higher number of eccentric LISA sources. These preliminary results
illustrate the importance of including weak interactions for accurately modeling how BBHs merge in clusters,
and how to link their emitted GW signals to their astrophysical environment.

\end{abstract}

\maketitle

\section{Introduction}\label{sec:Introduction}

The origin of binary black hole (BBH) mergers is still unknown, despite several now have been directly observed through their
emission of gravitational waves (GWs) \citep{2016PhRvL.116f1102A, 2016PhRvL.116x1103A, 2016PhRvX...6d1015A,
2017PhRvL.118v1101A, 2017PhRvL.119n1101A, 2019arXiv190210331Z, 2019arXiv190407214V}.
The currently observed set shows great variety in both the BH mass spectrum, and BH spins \cite[e.g.][]{2018arXiv181112907T}.
At least one neutron star (NS) merger has also been observed \citep{2017PhRvL.119p1101A} in addition to several
peculiar low signal-to-noise events, and this variety therefore indicates that
several formation channels might be at play. Some of the recently proposed channels include:
field binaries \citep{2012ApJ...759...52D, 2013ApJ...779...72D, 2015ApJ...806..263D, 2016ApJ...819..108B,
2016Natur.534..512B, 2017ApJ...836...39S, 2017ApJ...845..173M, 2018ApJ...863....7R, 2018ApJ...862L...3S},
dense stellar clusters \citep{2000ApJ...528L..17P,
2010MNRAS.402..371B, 2013MNRAS.435.1358T, 2014MNRAS.440.2714B,
2015PhRvL.115e1101R, 2016PhRvD..93h4029R, 2016ApJ...824L...8R,
2016ApJ...824L...8R, 2017MNRAS.464L..36A, 2017MNRAS.469.4665P},
active galactic nuclei (AGN) discs \citep{2017ApJ...835..165B,  2017MNRAS.464..946S, 2017arXiv170207818M},
galactic nuclei (GN) \citep{2009MNRAS.395.2127O, 2015MNRAS.448..754H,
2016ApJ...828...77V, 2016ApJ...831..187A, 2016MNRAS.460.3494S, 2017arXiv170609896H, 2018ApJ...865....2H},
very massive stellar mergers \citep{Loeb:2016, Woosley:2016, Janiuk+2017, DOrazioLoeb:2017},
and single-single GW captures of primordial black holes \citep{2016PhRvL.116t1301B, 2016PhRvD..94h4013C,
2016PhRvL.117f1101S, 2016PhRvD..94h3504C}.

The key exercise is how to observationally tell these channels apart; several observables have been suggested. For example,
an electromagnetic (EM) signal might be associated with both BBH mergers in AGN discs and those coming from rapidly rotating stars, in contrast to
gas-free isolated BBH mergers found in, e.g., stellar clusters and the field. Tidal disruption events (TDE) associated with
stars disrupted by BBHs in dense stellar clusters might also be used to probe the merger history of BBHs \citep{2019ApJ...877...56L, 2019arXiv190102889S, 2019arXiv190406353K}.
Other ways of `lightning up' single or binary BHs through TDEs include, e.g., BHs kicked into disruptive orbits from either dynamics, GW recoils or
direct collapse following supernova \citep[e.g.][]{2017PhRvL.118o1101S, 2019arXiv190509471F}.
Such multi-messenger (EM+GW) observables are interesting, but unfortunately also highly uncertain to both accurately predict and model.
More promising are properties that can be read off directly from the incoming
GW waveform, which include the BBH masses and spins, the orbital eccentricity, and even doppler effects related to a possible
movement of the BBH's center of mass (COM) \citep[e.g.][]{2017ApJ...834..200M, 2018arXiv180505335R}.
For example, the spins are expected to be isotropically distributed for BBH mergers forming in clusters, in contrast to field BBH
mergers \citep[e.g.][]{2016ApJ...832L...2R, 2017arXiv170601385F}. BBHs assembled in clusters
through chaotic few-body interactions have also recently been shown to lead to a non-negligible unique population of mergers with measurable
eccentricities in all bands from LISA (the Laser Interferometer Space Antenna) \citep{2018MNRAS.tmp.2223S, 2018MNRAS.481.4775D, 2019PhRvD..99f3003K} to LIGO (the Laser Interferometer Gravitational Wave Observatory) \citep{2006ApJ...640..156G, 2014ApJ...784...71S, 2017ApJ...840L..14S, 2017ApJ...840L..14S, 2018MNRAS.476.1548S, 2018ApJ...853..140S, 2019MNRAS.482...30S, 2018PhRvD..97j3014S, 2018ApJ...855..124S, 2019ApJ...871...91Z, 2018PhRvD..98l3005R}. Eccentric sources can form in other ways, e.g. through
Lidov-Kozai oscillations \citep[e.g.][]{2011ApJ...741...82T,2018ApJ...864..134R, 2018arXiv181110627F, 2019arXiv190208604R, 2019arXiv190309160F, 2019MNRAS.486.4443F, 2019arXiv190309659F}, quadruple systems \citep[e.g.][]{2019MNRAS.486.4781F}, and single-single GW
captures \citep[e.g.][]{2009MNRAS.395.2127O, Kocsis:2012ja, 2016PhRvD..94h4013C, 2018ApJ...860....5G}, but eccentric cluster
mergers are unavoidably produced if BHs are present in clusters, in particular globular clusters (GCs), and therefore constitute a highly useful diagnostic for
the cluster channel.

The topic of this paper is on how BBHs are dynamically driven to merger in GCs under the influence of both weak and strong interactions,
as well as what unique GW signals to expect from this population. BBHs in GCs generally
form through single-single-single interactions \citep[e.g.][]{Heggie:1975uy, 1976A&A....53..259A, Hut:1983js} in the cluster center, where the mass
segregated BH population constitute a highly dense BH subsystem \citep[e.g.][]{2018MNRAS.478.1844A, 2018MNRAS.479.4652A}. Each BBH
formed in this way is initially very wide and will not merge within a Hubble time, but subsequent interactions with incoming single BHs
lead to shrinkage of the orbit, referred to as hardening, and large changes in the eccentricity \citep[e.g.][]{Heggie:1975uy}. This eventually drives the BBH to merger
either in-side or out-side of its cluster. Current state-of-the-art Monte Carlo (MC) cluster codes aimed at modeling the evolution of
such BBHs, including the CMC code \citep[]{2018PhRvD..98l3005R} and the MOCCA code \citep[]{2013MNRAS.431.2184G},
incorporate only the effects from strong encounters,
and include therefore not the potential importance of the many more weak encounters a given BBH undergoes during its life inside its cluster.

In this work we present the first study on
how the numerous number of unbound weak encounters a BBH undergoes in-between each strong encounter affects its dynamical evolution and GW merger
properties. Each of these weak encounters changes the eccentricity of the BBH, which leads to a
non-linear and asymmetric random walk with varying step sizes, also known as a quasi-Levy flight,
in-between each strong encounter. In classical cluster dynamics where post-Newtonian (PN) corrections \citep[e.g.][]{2014LRR....17....2B} are
not included and all objects are assumed to be point-like, weak encounters are not important as the associated energy exchange is exponentially suppressed; however,
when dissipative effects, such as PN corrections, tides, and finite sizes are included, objects from ordinary stars to BHs can be driven to merger through
high eccentricity orbits \citep[e.g.][]{2017ApJ...846...36S}. Such eccentricity driven mergers not only give rise to interesting observables,
including GWs, TDEs, and jetted EM sources, but will also change the energy balance of the cluster and thereby its large scale dynamics.

We here explore the combined effects from strong and weak three-body encounters using analytical and numerical methods, and investigate how their interplay
drives BBHs to merger in dense clusters. For this we particularly make use of the strong scattering model presented in \citep{2018PhRvD..97j3014S} and \citep{2018MNRAS.tmp.2223S},
and our new second-order weak interaction solution presented in \citep{2019arXiv190409624H}, referred to as HS19, which expands upon
the first-order solution presented in \citep{1996MNRAS.282.1064H}, referred to as HR96.
We find that the inclusion of weak interactions leads to notable changes in both the number of
BBHs merging inside their cluster and their distribution across GW peak frequency and orbital eccentricity. This leads us to conclude that weak encounters are
important to include for an accurate modeling of how BBHs merge in clusters, which is naturally an increasingly relevant inquiry now that there is a growing population
of observed GW events. This poses some interesting challenges, as the leading numerical schemes which include the
H\'{e}non method in the field of PN dynamics do not easily allow for the inclusion of weak encounters.
We also highlight the importance of using the second-order results from HS19, which solve some crucial problems occurring when using the expressions of HR96 for evolving BBHs in the high eccentricity regime.

The paper is organized as follows. We start by describing the outcome of a single weak encounter in Section \ref{sec:A Single Weak Encounter}.
This includes a discussion of why distant encounters must be important for an accurate modeling of BBH mergers, and several new dynamical
effects that have not yet been properly described in the literature related to this subject. After this, we study in Section \ref{sec:Black Holes Binaries in Clusters}
a population of BBHs and how they evolve under the influence of both strong and weak encounters.
For this we use numerical and analytical techniques combined with MC methods.
We discuss and conclude our findings in Section \ref{sec:Conclusions}.

\section{A Single Weak Encounter}\label{sec:A Single Weak Encounter}

\begin{figure}
\centering
\includegraphics[width=\columnwidth]{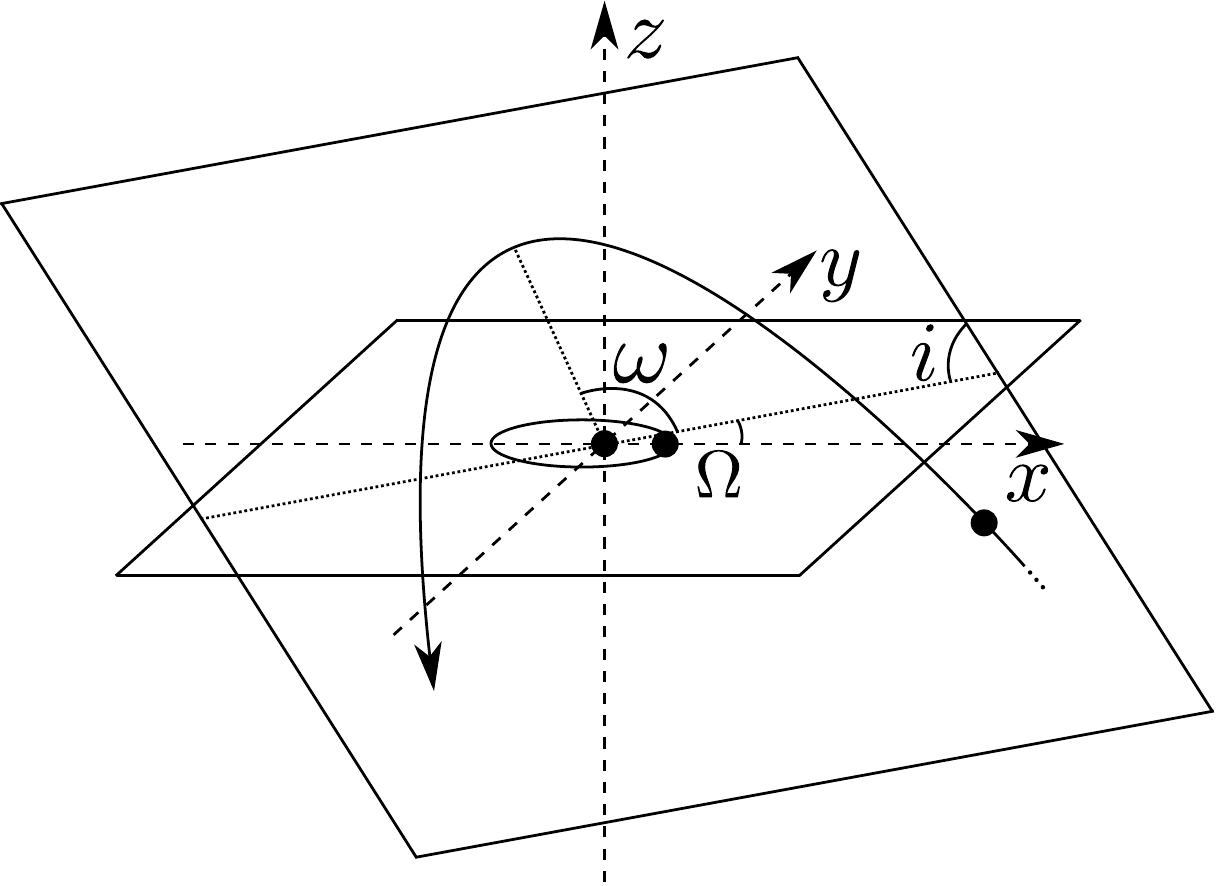}
\caption{Illustration showing how we define our orbital elements used throughout the paper. The binary is positioned in the $x,y$-plane
with its peri-center pointing along the $x$-axis, where the perturber is here illustrated to encounter the binary in its own unperturbed plane on a counter-clockwise
orbit. The black dots denoting the three interacting objects, and the solid lines showing their orbits, are not to scale and just added for illustrative purposes. 
We use `standard angles' defined as follows: $\Omega$ is the longitude of the ascending node measured in the plane of motion of the binary, $i$ is the inclination between
the two orbital planes, and $\omega$ is the longitude of peri-center of the third body measured in its plane of motion from the ascending node. Our definitions
are identical to the one used in HR96, where further details can be found. }
\label{fig:orbitill}
\end{figure}

We start by exploring the range of outcomes associated with a single weak encounter between a BBH and a single incoming BH.
For this, we consider a BBH with initial orbital eccentricity $e_{0}$ and semi-major axis (SMA) $a_{0}$ interacting with an incoming BH on a parabolic orbit.
Here, and in the rest of the paper, we assume the interacting BHs all have the same mass $m$. For describing the ICs we make use of the
standard orbital angles shown in Figure \ref{fig:orbitill}, which is further described in HR96. Note here that these are different from the set used in HS19 (see Section 2.1 in the latter paper).

\subsection{Background and Motivation}\label{sec:Background and Motivation}

For binary-single interactions to be in the secular regime, the mean motion of the binary must be much greater than
the orbital speed of the single at the time of peri-center passage (HR96). In this limit, the binary-single equations-of-motion (EOM) can be written out using
a technique known as `orbital averaging', which is widely used for e.g. studying the hierarchical bound-triple Lidov-Kozai problem \citep[e.g.][]{2016ARA&A..54..441N}.
In the case where the third object is on an unbound orbit with an initial eccentricity $E$, the orbital averaged
EOM naturally gives rise to the following parameter $\epsilon$, which appears as a common factor in the EOM (HS19),
\begin{equation}
\epsilon = \left[ \frac{M^{2}}{m_b(m_b+M)} \left( \frac{a_0}{r_{\rm p}} \right)^{3} (1+E)^{-3} \right]^{1/2},
\label{eq:epsilon}
\end{equation}
where $m_b$ is the mass of the BBH ($m_b=2m$), $M$ is the mass of the perturber ($M=m$), and $r_{\rm p}$ is the distance
between the BBH center-of-mass (COM) and the incoming BH at the time of peri-center passage.
As seen, weak interactions for which $r_{\rm p}/a{_0} \gg 1$ are characterized by having $\epsilon \ll 1$.

The main effect from a BBH undergoing a weak interaction with a single incoming BH
is a change in the orbital eccentricity of the BBH, $\Delta{e}$, such that the final eccentricity after the encounter is $e_{\rm fin} = e_{0} + \de$.
The change $\Delta{e}$ can to first-order (FO) in $\epsilon$ in the parabolic limit be analytically expressed as (HR96),
\begin{equation}
\de_{\rm FO} = - \epsilon \frac{15\pi}{4}e_0\sqrt{1-e_0^2}\sin{2\Omega}\sin^{2}{i},
\label{eq:de1ord}
\end{equation}
where $\Omega$ is the longitude of the ascending node measured in the plane of motion of the binary, and $i$ is the inclination between
the two orbital planes of the binary and the perturber, respectively, as further described in Fig. \ref{fig:orbitill}.
Note here that this term has been derived by expanding the EOM to
lowest non-trivial order in the (assumed to be) small ratio of the binary separation to the distance between the binary COM and the perturber, i.e., it is only accurate to quadrupole order.
As shown in HR96, this expression accurately fits numerical simulations
for $r_{\rm p}/a_0 \gg 1$ and intermediate values of $e_0$, and have therefore been widely used as a replacement for otherwise
computationally demanding three-body simulations \citep[e.g.][]{2019ApJ...872..165G}. HR96 also presented solutions to the $e_0=0$ limit, and in regions where
the secular approach breaks down.
However, one consideration not addressed in HR96 is the regime where $\de \approx 1-e_0$, which is possible if $e_{0}$ is near unity. 
For describing the formation of BBH mergers in dense clusters this is a highly relevant limit, as basically all mergers form through high eccentricity orbits.
It is known that BBHs can be driven to very high eccentricities entirely through strong few-body encounters, i.e. without any
assistance from weak interactions (see e.g. \citep{2018PhRvD..97j3014S}); however, binaries
undergoing strong interactions will also undergo numerous weak interactions in-between each strong interaction each of
which will perturb the binary \citep[e.g.][]{2019ApJ...872..165G}.
The question is if a model that also includes weak interactions will lead to more or less mergers from high eccentricity orbits.

\begin{figure}
\centering
\includegraphics[width=\columnwidth]{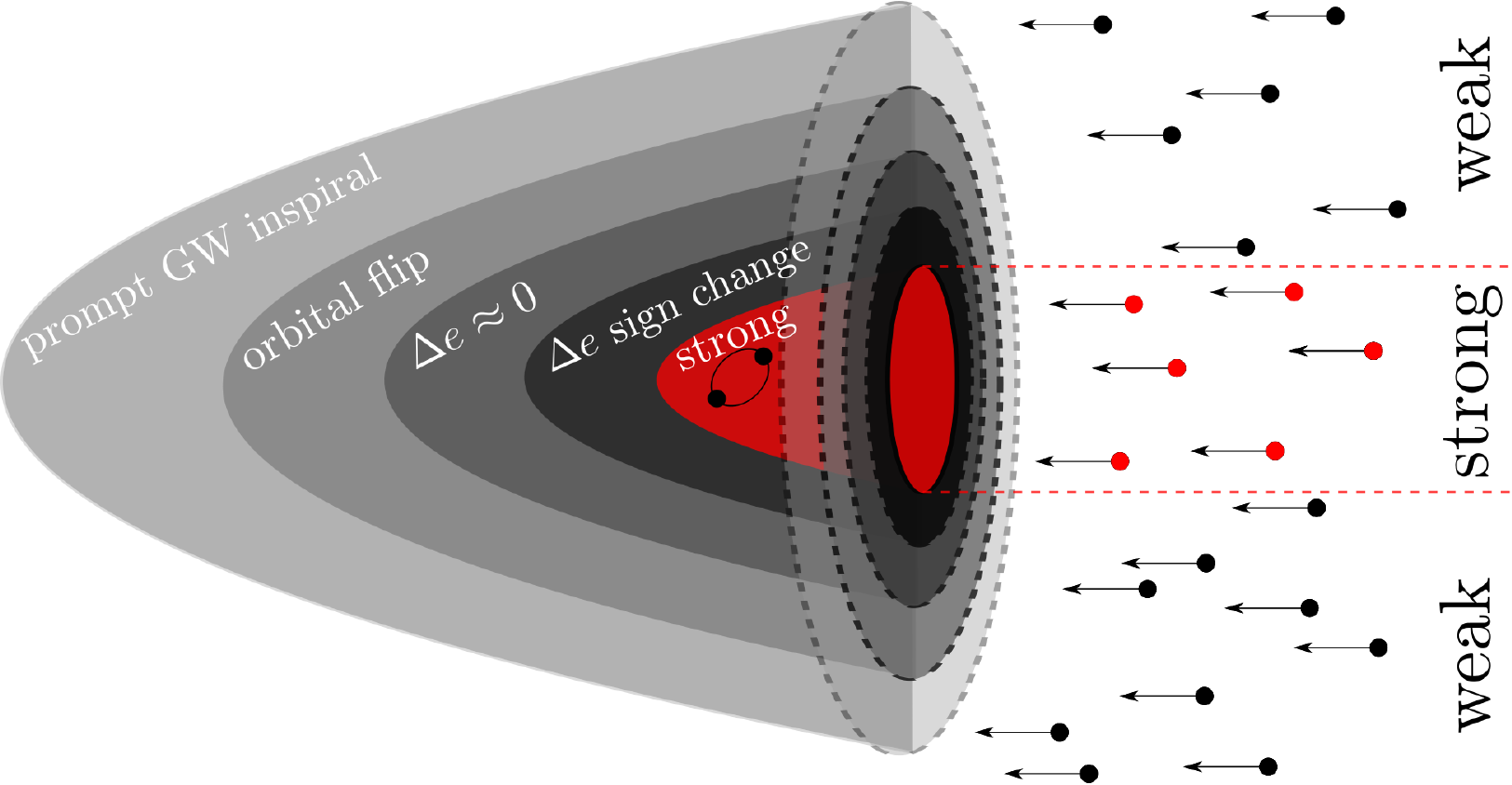}
\caption{Illustration of a binary interacting with an incoming population of
single objects, where the strong interaction regime is highlighted in {\it red}, and the weak interaction regime in {\it grey}.
The weak regime is further divided into different sub-regimes, each giving rise to a distinct type of outcome.
The range of outcome types depends on the ICs, the shown here are the ones
relevant for our fiducial parameters outlined in the end of Section \ref{sec:Background and Motivation}.
As described in Section \ref{sec:A Single Weak Encounter}, the outcomes of a binary-single interaction with our chosen ICs
can, from high to low values of $r_{\rm p}/a_0$, broadly be divided into the following regimes:
$(i)$ Distant weak interaction regime. Here the change in binary eccentricity is relatively small,
and can be described using FO theory. For our ICs $\de > 0$.
$(ii)$ Prompt GW inspiral regime. Here the final eccentricity is $\efin \approx 0$,
which results in a prompt GW inspiral merger during the interaction. This also marks the limit below which SO theory has to be used.
$(iii)$ Orbital flip regime. Below this limit the orbital angular momentum vector flips around during the interaction.
$(iv)$ $\de \approx 0$ regime. The final eccentricity is here $\efin \approx e_0$, implying $\de \approx 0$.
$(v)$ $\de$ sign-change regime. After crossing $\efin = e_0$ from above the binary here ends up with $\efin < e_0$, i.e. $\de < 0$.
$(vi)$ Strong regime. The binary here undergoes a strong interaction with an outcome that can be described using probability theory.
In Section \ref{sec:Black Holes Binaries in Clusters} we consider the evolution of a BBH undergoing strong and weak interactions
inside a dense cluster.
}
\label{fig:bs_ill}
\end{figure}

To gain insight into this and motivate our present study, we now consider the following
simple cluster model: we consider a cluster core described by a BH subsystem with a constant density and velocity dispersion,
and assume that all BH interactions take place in this system and only involve objects of similar mass.
This model provides in fact a good description of real cluster cores, as BHs both are expected to segregate, separate in mass, and clear out
the center, as a result of mass segregation. In this picture, if only strong binary-single encounters
are included in the cluster dynamics, a BBH will merge in-between its interactions if its GW inspiral life time \citep{Peters:1964bc},
\begin{equation}
\tau \approx \frac{768}{425}\frac{5c^{5}}{512G^{3}} \frac{a^{4}}{m^{3}}\left(1- e^2 \right)^{7/2},
\label{eq:tGW}
\end{equation}
is less than its strong encounter (SE) time, $t_{\rm SE}$, defined as the time it takes for the BBH
to undergo its next strong interaction \citep[e.g.][]{2018MNRAS.tmp.2223S},
\begin{equation}
t_{\rm SE} \approx  \frac{1}{12 \pi G} \frac{v_{\rm dis}}{n_{\rm s} m a},
\label{eq:tSE}
\end{equation}
where $n_{\rm s}$ is the number density of single BHs in the cluster, $v_{\rm dis}$ is the cluster velocity dispersion, $a$ is the
SMA of the BBH, and $e$ is its eccentricity. This correspondingly implies that a BBH with an eccentricity $e > e_{\rm SE}$,
\begin{equation}
e_{\rm SE} \approx \sqrt{1- \left(t_{\rm SE}/ \tau_{\rm c} \right)^{2/7}},
\label{eq:eGW}
\end{equation}
where $\tau_{\rm c}$ denotes the time for which $e=0$ in Eq. \eqref{eq:tGW},
will undergo a merger before its next strong binary-single interaction. 
For a BBH with $a = 1$ AU, $m = 20\ M_{\odot}$, $n_{\rm s} = 10^{5}\ \text{pc}^{-3}$,
and $v_{\rm dis} = 10\ \text{kms}^{-1}$, one finds $e_{\rm SE} \approx 0.99$. This is a relatively high eccentricity,
which means that even small changes of order $\de \sim 1-0.99 \sim 10^{-2}$ can change the state and outcome
of the BBH, from e.g. surviving to merging. Now, changes of such magnitude can easily be provided by weak interactions;
the question is then, will weak interactions on average prevent or help a BBH to merge if its eccentricity is already near the critical
value $e_{\rm SE}$? As pointed out above, the FO result $\de_{\rm FO}$ presented in HR96 breaks down in this high
eccentricity limit, and will therefore often lead to unphysical results such that $e_\mathrm{fin} \geq 1$.

\begin{figure}
\centering
\includegraphics[width=\columnwidth]{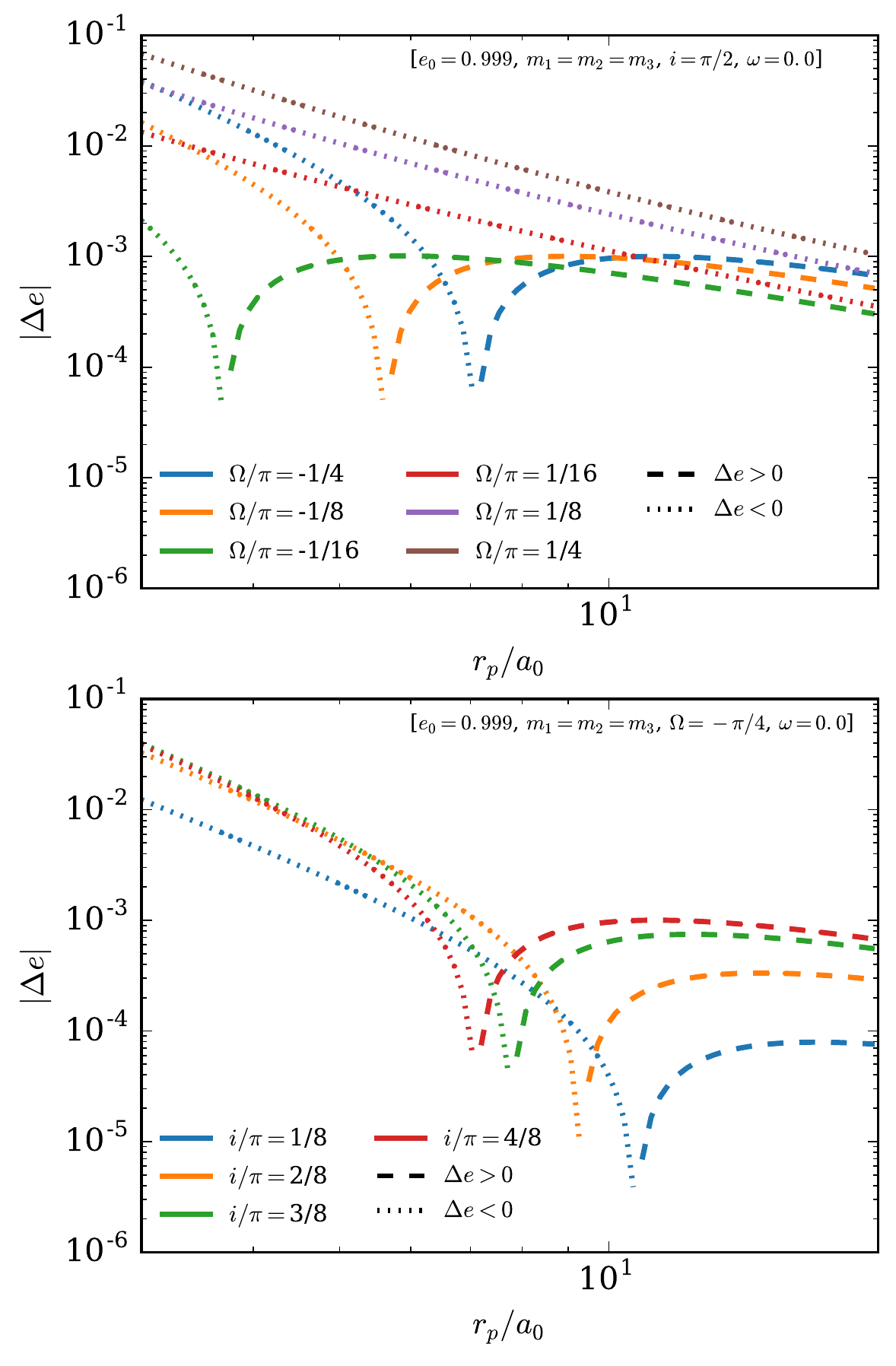}
\caption{Change in eccentricity, $\de$, derived using our SO solution given by Eq. \eqref{eq:de2ord}.
The FO solution given by Eq. \eqref{eq:de1ord}, always predicts a simple power-law change
of the form $\de \propto (r_{\rm p}/a_0)^{-3/2}$. The large variations seen above are therefore due to corrections from the
SO term $\propto \epsilon^2$.
In the above plots, the {\it dashed} and {\it dotted} lines show the change in eccentricity for $\de > 0$ and $\de < 0$, respectively,
where the different {\it colored} lines show results for different values of the orbital angles, as specified in the legends. The initial values
that are kept fixed are shown in the upper right corner in each plot. As seen, in the {\it top plot} the inclination $i = \pi/2$ and $\Omega$ is varied,
where in the {\it bottom plot} the angle $\Omega = -\pi/4$ and the inclination $i$ is varied. The SO term gives rise to very important and
easily notable corrections that greatly differ from the FO
prediction, which includes sign changes, orbital spin flips, and a different power-law scaling for very close encounters.
This is further described in Section \ref{sec:A Single Weak Encounter}.
A comparison to full numerical scatterings is shown in Fig. \ref{fig:de_etc_rp} and HS19.
Finally, one should note that when 1,2PN are included any reference to initial orbital elements is ill-defined
due to precession. For this reason, we do not include lower PN terms for this and similar figures, which are aimed at giving a clear overview of
outcomes. In reality, precession will to `first order' just add to a randomization of the initial orbital elements, which is naturally taken into account
in standard MC experiments.
}
\label{fig:deHS19_ill}
\end{figure}

To resolve this problem with the FO solution in the high eccentricity limit, we used in our companion paper HS19 second-order (SO) perturbation theory
to derive higher order correction terms to $\de$. In short, in the FO approximation of HR96, the binary orbital components are assumed to be fixed while integrating 
over the passage of the perturber (this is analogous to `double averaging' in hierarchical triple systems). Instead, in HS19 we used Fourier expansions to take into 
account the instantaneous response of the binary to the perturber while integrating over the latter's passage. This resulted in an expression for the eccentricity change to SO in $\epsilon$, which in the limit of small eccentricity changes and parabolic perturbers reads,
\begin{align}
\label{eq:de2ord}
\de_{\rm SO} =	& \de_{\rm FO} + \epsilon^2 \frac{3}{512} \pi e_0 \biggl [ - 100 \left(1-e_0^2\right) \sin 2 \Omega \\
\nonumber	& \biggl \{ \left (5 \cos i+3 \cos 3 i\right ) \cos 2 \omega +6 \sin i \sin 2 i\biggl \} \\
\nonumber	& +4 \cos 2 i \biggl \{3 \pi  \left(81 e_0^2-56\right) - 200 \left(1-e_0^2\right) \\
\nonumber	& \cos 2 \Omega  \sin 2 \omega \biggl \} + 3 \pi  \biggl \{ 200 e_0^2 \sin ^4i \cos 4 \Omega \\
\nonumber	& +8 \left(16 e_0^2+9\right) \sin ^2 2 i \cos 2 \Omega  \\
\nonumber	& + \left(39 e_0^2+36\right) \cos 4 i - 299 e_0^2+124\biggl \}\biggl ],
\end{align}
where the introduced set of angles $\{i, \Omega, \omega \}$ are shown and defined in Fig. \ref{fig:orbitill}. We reiterate that the orbital angles used here are defined differently than in HS19. 
As for $\de_{\rm FO}$, this term is only accurate to quadrupole order.
The second term that is SO in $\epsilon$, resolves many problems associated with only using the FO term $\de_{\rm FO}$,
as further explained in HS19. Note here that all dependencies on mass and closest approach only enter via the $\epsilon$ term.
A few examples of $\de_{\rm SO}$, as a function of the peri-center distance of the perturber w.r.t to the COM of the BBH
scaled by the initial SMA, $r_{\rm p}/a_0$, are shown and discussed in Fig. \ref{fig:deHS19_ill}.
As seen, the SO correction clearly breaks the simple FO power-law solution. 
This will be described in greater detail later.
Finally, one should note that the SO solution is still only accurate to quadrupole order and will therefore
break down when $r_{\rm p}/a_0$ starts to approach unity, and/or when the binary's component masses are unequal. In contrast, the FO solution can break down at {\it any} $r_{\rm p}/a_0$, i.e. also in highly secular regions,
depending on the initial eccentricity. The improvements from including octupole terms (the next non-trivial-order terms in the expansion of the small ratio of the binary separation to the distance between the binary COM and the perturber), as well has higher-order terms in $\epsilon$ (i.e,. third order), will be explored in upcoming work.

In the subsections below we systematically describe the outcomes of a BBH undergoing a single weak encounter, and the associated novel features that
are not captured by the classical FO result $\de_{\rm FO}$ from HR96. All our results are based on the initial set of orbital angles
$ i = \pi/2$, $\Omega = -\pi/4$, $\omega = 0$, a set we loosely refer to as our
`fiducial parameters',  to match previous results from HR96 for easy comparison.
We therefore here focus on the dependence on $r_{\rm p}/a$, but we will in an upcoming paper
present further discussions on the dependence on the initial set of orbital angles.
An illustration outlining the different outcomes associated with our choice of fiducial parameters is shown in Fig. \ref{fig:bs_ill}.

\subsection{Distant weak encounters}\label{sec:Distant weak encounters}

\begin{figure}
\centering
\includegraphics[width=\columnwidth]{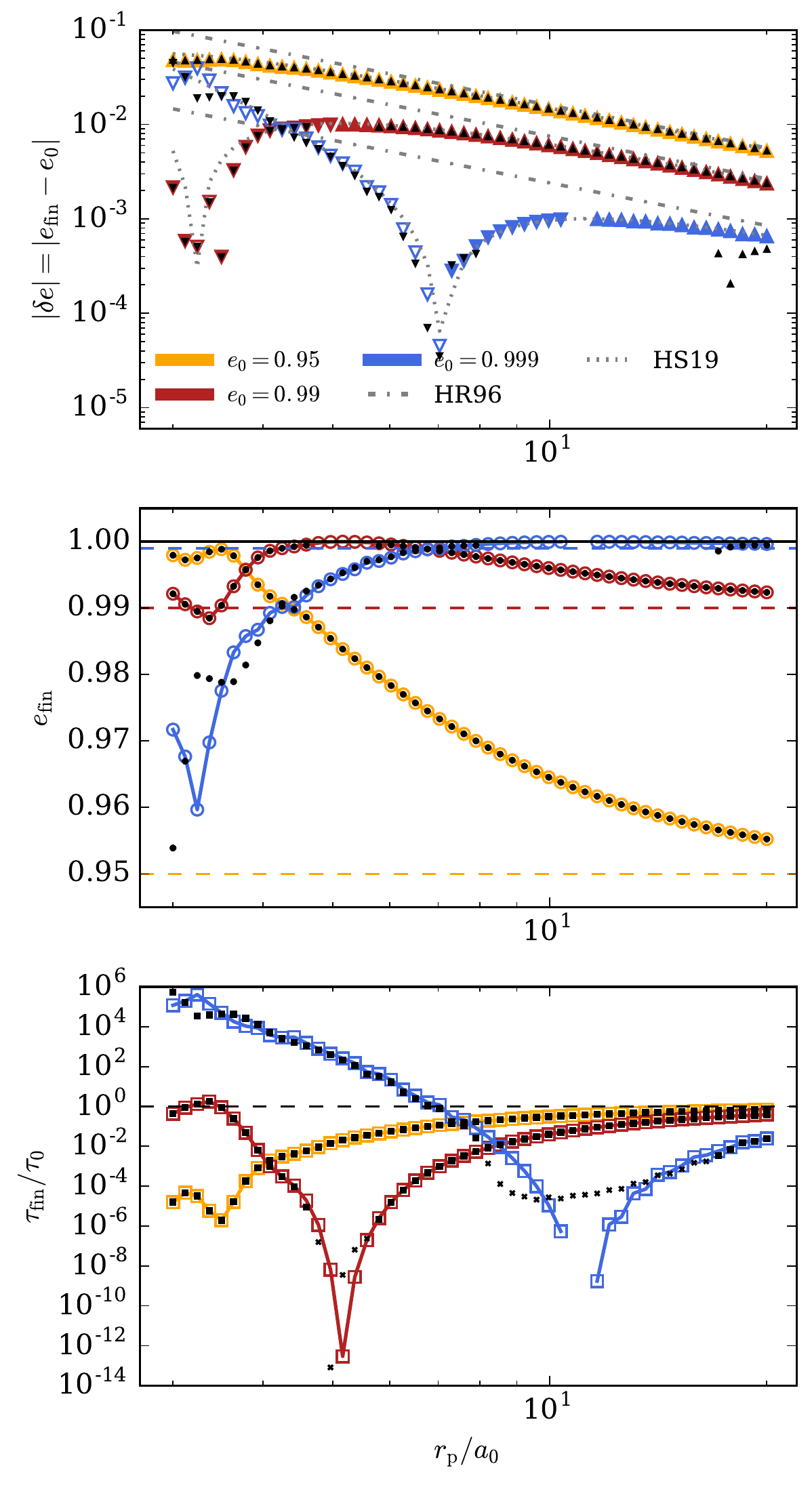}
\caption{Results from weak binary-single interactions between three equal mass BHs for varying values of $e_0$ ({\it orange/red/blue}) and
$r_{\rm p}/a_0$ (x-axis). For all scatterings we assume: $i = \pi/2$, $\Omega = -\pi/4$, $\omega = 0$, $m = 20M_{\odot}$, and $a_0 = 0.5$ AU.
The {\it plot symbols} denote results from full numerical three-body integrations,
where the larger {\it colored symbols} show results from including Newtonian forces only, and the smaller {\it black symbols} show results from additionally
including the 2.5PN term.
{\it Top plot}: Change in eccentricity $|\de|$. The {\it full} and {\it empty} symbols denote $\de > 0$ and $\de < 0$, respectively, where
an upwards (downwards) pointing {\it triangle symbol} denotes $\text{sign}(J_{\rm BH z}) = +1 (-1)$.
The {\it grey dashed-dotted} lines show predictions from the FO solution given by
Eq. \eqref{eq:de1ord} (HR96), where the {\it grey dotted} lines show predictions from our SO result given by \eqref{eq:de2ord} (HS19).
{\it Middle plot}: Final eccentricity $e_{\rm fin}$. The {\it dashed lines} are here showing the initial values $e_0$.
{\it Bottom plot}: GW inspiral time of the binary after the interaction, $\tau_{\rm fin}$, scaled by the initial value $\tau_0$.
The {\it black crosses} denote outcomes for which the BBH merges during the weak interaction.
}
\label{fig:de_etc_rp}
\end{figure}

We start by considering the top plot in Fig. \ref{fig:de_etc_rp}, which shows $|\de| = |\efin - e_0|$ as a function of
$r_{\rm p}/a_0$. The dashed-dotted lines show the FO result $\de_{\rm FO}$ from HR96 (Eq. \eqref{eq:de1ord}), 
the dotted lines show the SO result $\de_{\rm SO}$ from HS19 (Eq. \eqref{eq:de2ord}), where the
symbols show results from a full three-body integration using the $N$-body
code used in \citep{2017ApJ...846...36S, 2018ApJ...855..124S, 2018MNRAS.481.5436S}. Note here that only one simulation is
performed per point, which naturally gives rise to Poisson scatter.

The numerical and the analytical results clearly approach the simple power-law solution
$\de \propto (r_{\rm p}/a_0)^{-3/2}$ given by Eq. \eqref{eq:de1ord} in the asymptotic limit $r_{\rm p}\gg a$. The slope value $-3/2$ is universal in the
sense that it does not depend on the mass ratio or the angular configurations of the three-body system.
This universality provides some insight into the role of how important a constant flux of distant encounters might be for the
eccentricity evolution. Especially, we might ask the question: does a uniform and infinite sea of weak encounters lead to a
divergent result? To gain some insight into this we first notice that the number of single encounters $dN_{\rm s}$ per
time interval $dt$, $\Gamma_{\rm s}$, with $r_{\rm p} < R_{\rm p}$ under the assumption of constant number density and
velocity dispersion is $\propto R_{\rm p}$ \citep{2018ApJ...853..140S}.
This implies that the rate per peri-center interval $d\Gamma_{\rm s}/dr_{\rm p} \propto k$, where $k$ is a constant. Let us now consider
a `worst case scenario' where all single encounters lead to the same positive change in eccentricity $\de_{+} \propto (r_{\rm p}/a_0)^{-3/2}$
(note that $\langle \de \rangle = 0$ to FO, but not the variance).
Assuming all the eccentricity changes are uncorrelated, one finds in this case that the total change in eccentricity, $d\de_{\rm tot}$, per unit time, $dt$,
from single encounters in the range
$r_{\rm p,min}$ to $r_{\rm p,max}$ is given by
\begin{equation}
\frac{d\de_{\rm tot}}{dt} \propto \int_{r_{\rm p,min}}^{r_{\rm p,max}} \frac{d\Gamma_{\rm s}}{dr_{\rm p}} \de_{+} dr_{\rm p} \propto \frac{1}{\sqrt{r_{\rm p,min}}} - \frac{1}{\sqrt{r_{\rm p,max}}}.
\label{eq:eGW}
\end{equation}
This term does not diverge when setting the upper limit $r_{\rm p,max} = \infty$, which illustrates
that the rate of eccentricity change from even an infinite number of distant encounters all leading to
a positive change $\de_{+} \propto (r_{\rm p}/a_0)^{-3/2}$, does not lead to
a diverging result. We therefore expect our results to converge as the upper value of $r_{\rm p}$ in our analysis
increases; a value we generally will refer to as $R_{\rm p}$. We note here that an accurate analytical modeling of how the eccentricity changes over time
from close and distant encounters can be done using a Boltzmann-approach, which requires the derivation of
diffusion coefficients. We will do that in an upcoming paper.
In this paper we will be using numerical MC techniques, as described later in Section \ref{sec:Black Holes Binaries in Clusters}.

Finally, the discussions from this section are based on a Newtonian treatment of the problem; however,
when PN terms are included several changes are expected from especially precession, which will
make the ICs ill-defined and also modify the simple asymptotic solution $\de \propto (r_{\rm p}/a_0)^{-3/2}$.
This will be discussed further in Section \ref{sec:Post-Newtonian Effects}.

\subsection{Orbital flip and prompt GW Inspiral}\label{sec:Orbital flip and prompt GW Inspiral}

We now consider lower values of $r_{\rm p}/a_0$ where the FO power law solution $\de \propto (r_{\rm p}/a_0)^{-3/2}$ from Eq. \eqref{eq:de1ord} starts to break down.
Considering our examples, we see in the top plot of Fig. \ref{fig:de_etc_rp} that for BBHs with an initial eccentricity $e_0 = 0.95$, $e_0 = 0.99$,
and $e_0 = 0.999$ that the FO estimates start to deviate from the
numerically estimated (correct) solutions for $r_{\rm p}/a_0 \lesssim 5$, $r_{\rm p}/a_0 \lesssim 7$, and $r_{\rm p}/a_0 \lesssim 20$, respectively.
Generally, the value for $r_{\rm p}/a_0$ below which the FO result starts to break
down, a value we denote $(r_{\rm p}/a_0)_{\rm FO}^{\rm BR}$, is where the change in eccentricity $\de$ is comparable to
$1-e_0$, i.e. where the final eccentricity $\efin \approx 1$. The physical boundary $e = 1$ for bound orbits is not built into the
the FO solution, therefore the decrease seen in the figure for the numerically estimated value of $\de$ compared to the FO power law prediction
for $(r_{\rm p}/a_0) \lesssim (r_{\rm p}/a_0)_{\rm FO}^{\rm BR}$ originates to simply `prevent' the BBH for reaching values of $\efin > 1$.
This statement is further supported by considering the middle plot of Fig. \ref{fig:de_etc_rp}, which shows $\efin$ as a function of $r_{\rm p}/a_0$.

\begin{figure}
\centering
\includegraphics[width=\columnwidth]{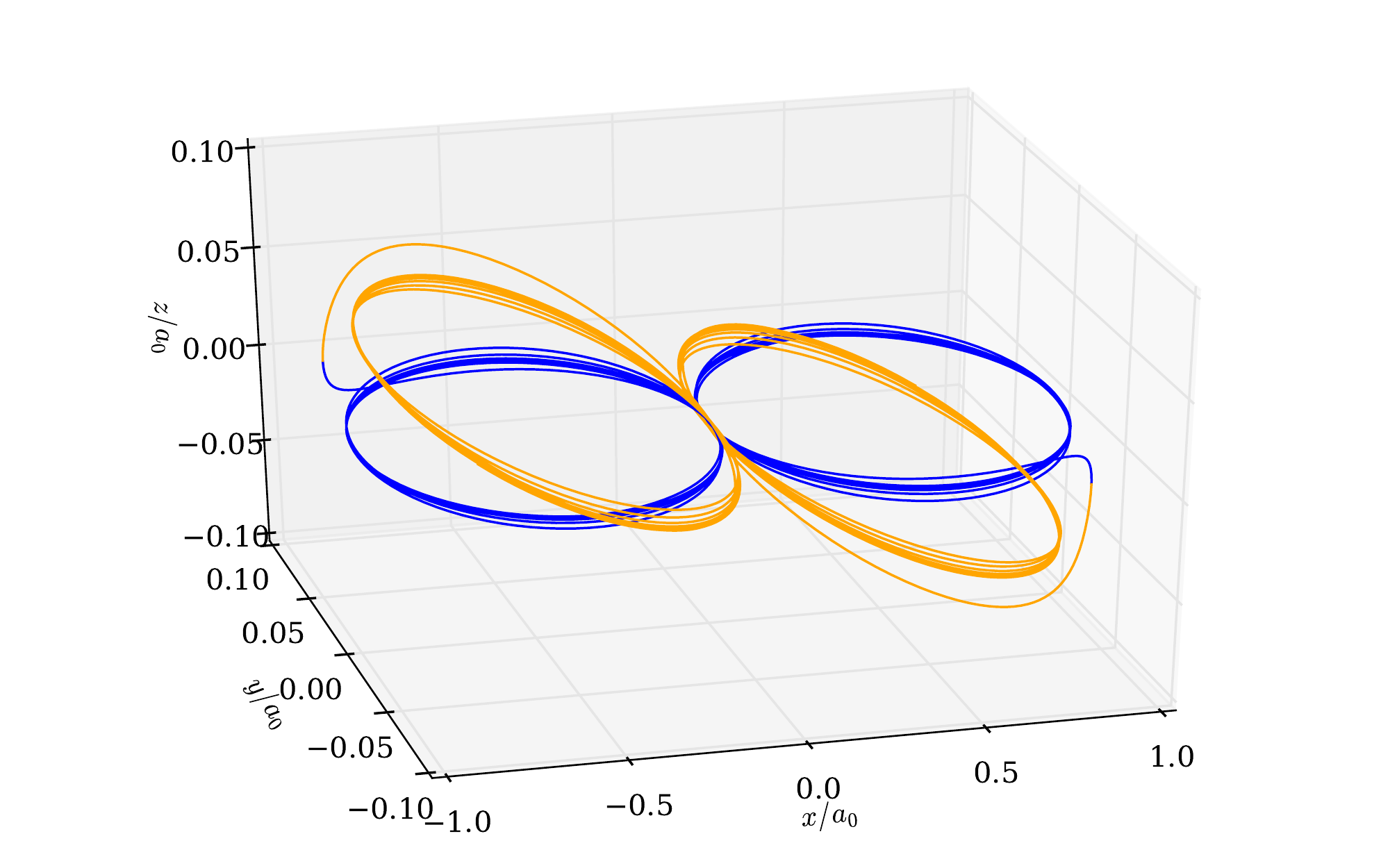}
\includegraphics[width=\columnwidth]{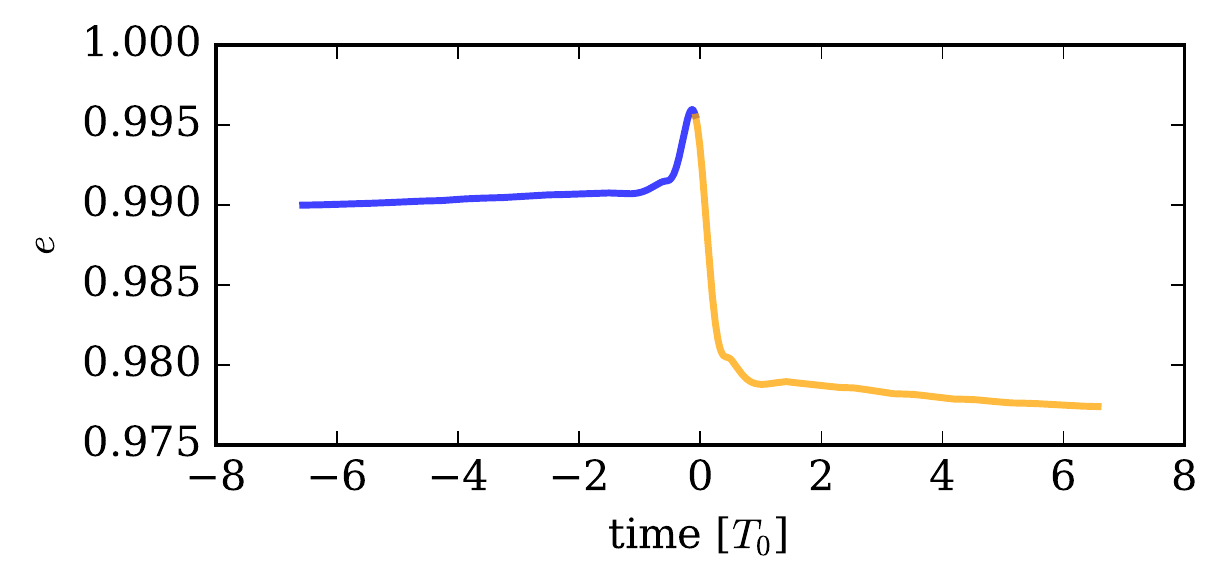}
\includegraphics[width=\columnwidth]{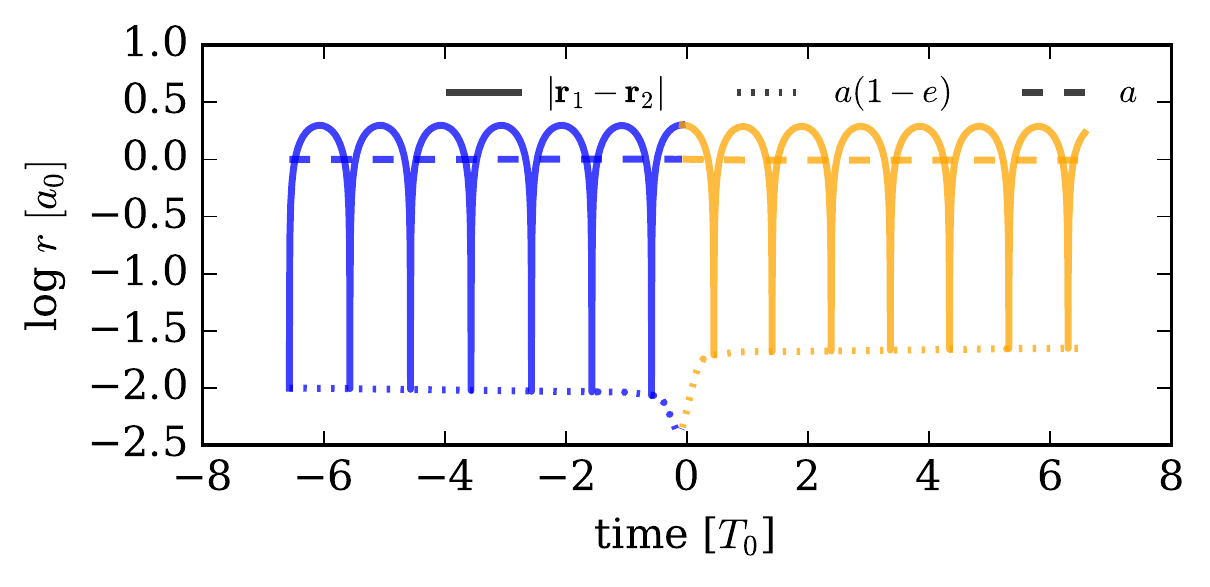}
\caption{Binary evolution during a single weak encounter with initial orbital parameters: $r_{\rm p}/a_0 = 3.0$, $e_0 = 0.99$,
and $i = \pi/2$, $\Omega = -\pi/4$, $\omega = 0$. The {\it line colors} denote the sign of the BBH orbital angular momentum vector
along the $z$-axis, where blue (orange) shows the part of the orbit where $\text{sign}(J_{\rm BH z}) = +1$ ($\text{sign}(J_{\rm BH z}) = -1$).
The initial $J_{\rm BH z}$ is aligned with the $z$-axis, and the change from blue to orange therefore illustrates how the orbit
`flips around'. The simulations do not include PN effects.
{\it Top plot}: Illustration of the actual BBH orbits printed out from our three-body integrator. As seen, the encounter in this example courses the BBH orbit to flip on
a time scale shorter than the BBH orbital time.
{\it Middle plot}: Corresponding evolution of the BBH orbital eccentricity calculated from its osculating elements.
In FO theory $\de$ would be so large that $\efin > 1$ (see e.g. Fig. \ref{fig:de_etc_rp}). However, this is not what physically happens.
Instead, the BBH first approaches $e \sim 1$ after which
it flips around and continue its evolution on a counter rotating orbit.
{\it Bottom plot}: Distance, $|\bf{r}_{1} - \bf{r}_{2}|$, osculating peri-center distance, $a(1-e)$, and SMA, $a$, of the BBH as a function of time.
From the osculating peri-center distance one would naively conclude that the two BHs undergo a very close passage and
possible merger during the orbital flip; however, as the flip happens almost instantaneously, the two BHs never get particular close as
seen when comparing to $|\bf{r}_{1} - \bf{r}_{2}|$. These results are further discussed in Section \ref{sec:Orbital flip and prompt GW Inspiral}
and \ref{sec:Sign Reversal}.
}
\label{fig:3bodysimJflip}
\end{figure}

An approximate value for $(r_{\rm p}/a_0)_{\rm FO}^{\rm BR}$ can be found by considering the relation $\de = 1-e_0 = \de_{\rm FO}$
and then solve for $r_{\rm p}/a_0$. In the parabolic equal mass limit one finds,
\begin{equation}
(r_{\rm p}/a_0)_{\rm FO}^{\rm BR} \approx \left[ \frac{e_0\sqrt{1-e_0^2}}{e_0-1} \frac{15 \pi \sin{2\Omega} \sin^{2}{i}}{16\sqrt{3}} \right]^{2/3}.
\label{eq:rpbreak}
\end{equation}
For our `fiducial orbital parameters' listed in Section \ref{sec:Background and Motivation} and for $e_{0} = 0.99$,
one finds using the above relation $(r_{\rm p}/a_0)_{\rm FO}^{\rm BR} \approx 8$, which indeed provides a reasonable estimate
for when FO theory starts to break down. In the limit of very high initial eccentricity one finds that
$(r_{\rm p}/a_0)_{\rm FO}^{\rm BR} \propto (1-e_0)^{-1/3}$, which shows that for $e_{0} = 1-10^{-x}$ the
break value $(r_{\rm p}/a_0)_{\rm FO}^{\rm BR} \propto 10^{x/3}$. One decade closer to $e=1$ therefore moves the break value
up by about a factor of $2$.
Finally, one should note that for decreasing values of $e_0$ the value of $(r_{\rm p}/a_0)_{\rm FO}^{\rm BR}$
correspondingly approaches $1$ in which case the secular formalism breaks down. The above relation is therefore only useful for BBHs
with an initial high eccentricity.

We now focus on our SO expression $\de_{\rm SO}$ given by Eq. \eqref{eq:de2ord}, which is shown with
dotted lines in the top plot of Fig. \ref{fig:de_etc_rp}. As seen here and further discussed in HS19, this relation accurately captures
the correct physical form for $\de$ and thereby resolves the problems associated with the FO result near and below 
$(r_{\rm p}/a_0)_{\rm FO}^{\rm BR}$. The decrease in $\de$ given by the SO relation that ensures that $\efin < 1$ originates from the
second term $\propto \epsilon^{2}$ in Eq. \eqref{eq:de2ord}, which in this case has a sign opposite to the FO term $\propto \epsilon$.
As shown in HS19, the location of $r_{\rm p}/a_0$ for which $\de$ is maximized, i.e. the location of the plateau seen in the
plot of $\de$ slightly below $(r_{\rm p}/a_0)_{\rm FO}^{\rm BR}$, can be analytically calculated. As in HS19, we refer to
the $(r_{\rm p}/a_0)$ location of the plateau by $(r_{\rm p}/a_0)_{\rm plateau}$. The solution for $(r_{\rm p}/a_0)_{\rm plateau}$ follows
from solving the equation $\partial \de_{\rm SO}/\partial(r_{\rm p}/a_0) = 0$, where $\de_{\rm SO}$ is given by Eq. \eqref{eq:de2ord}.
The solution is long, but can easily be written out in closed form, which provides a clear expression for
when SO terms are important (See HS19 for further details).

The dynamics near the plateau are highly interesting, and we now explore them in greater detail.
First, we find that as one passes the plateau from high to low values of $r_{\rm p}/a_0$,
the component of the BBH orbital angular momentum vector projected along the initial `z-axis' (perpendicular to the orbital plane of the BBH),
denoted here by $J_{\rm BH z}$, changes sign; a behavior we denote {\it orbital flipping}.
Loosely speaking, the vector component $J_{\rm BH z}$ flips around, from $\text{sign}(J_{\rm BH z}) = +1$ to $\text{sign}(J_{\rm BH z}) = -1$,
as the BBH `reflects off' the $e = 1$ boundary near $(r_{\rm p}/a_0)_{\rm plateau}$. The final BBH therefore spins in the opposite direction compared to its
orbital spin vector prior to the interaction. The sign change is also shown in the upper plot of Fig. \ref{fig:de_etc_rp}, where the triangles pointing upwards indicate
no sign change ($\text{sign}(J_{\rm BH z}) = +1$ ) and the triangles pointing downwards indicate a change ($\text{sign}(J_{\rm BH z}) = -1$). As seen,
the change indeed happens right near $(r_{\rm p}/a_0)_{\rm plateau}$ where $\efin \approx 1$.
We note here that a similar orbital-flip effect has been observed in the Lidov-Kozai case \citep[e.g.][]{2014ApJ...785..116L, 2018CeMDA.130....4S}.
To get a better view of the dynamics, we now consider the top plot in Fig. \ref{fig:3bodysimJflip}, which shows
how the flip occurs using a full three-body integration of the problem. The solid lines show the orbits of the two BHs in their COM,
where blue lines highlight their evolution before the orbital flip ($\text{sign}(J_{\rm BH z}) = +1$) and the orange lines
after the flip ($\text{sign}(J_{\rm BH z}) = -1$). As seen, the flip occurs on a timescale that is less than the
initial orbital time of the BBH, denoted by $T_{0}$, and therefore represents a `non--adiabatic maneuver'.
This leads us to another important point, namely that even though the BBH changes orbital spin direction,
it does not generally lead to a prompt merger during the flip, although the BBH angular momentum is able to vanish right in the flip (depending on the orbital
configurations). In other words, the so-called osculating elements of the BBH $a,e,r_{\rm BH} = a(1-e)$ might predict that $r_{\rm BH} \approx 0$ during the flip which would lead to
a prompt merger; however, as the flip does not happen adiabatically, but instead almost promptly in less than one orbit, the BBH is most likely never to pass through
the peri-center passage $r_{\rm BH} \approx 0$ as otherwise suggested by the osculating elements. This point is illustrated in the bottom plot of Fig. \ref{fig:3bodysimJflip},
which shows the actual distance between the two BHs in the binary, $|\bf{r}_1 - \bf{r}_2|$, and the osculating peri-center distance, $r_{\rm BH} = a(1-e)$, as a function
of time. As seen, right at the flip near $t/T_0 \approx 0$, the osculating value $r_{\rm BH} = a(1-e)$ suggests the two BHs (or stars) will undergo an extremely close passage
that might lead to a merger; however, the real distance $|\bf{r}_1 - \bf{r}_2|$ shows that they (in this example) never get any closer than their initial peri-center distance.
Therefore, one has to be very careful when using orbital average quantities, such as osculating elements, for estimating possible mergers during
an interaction.

\subsection{Sign Reversal}\label{sec:Sign Reversal}

The last classical effect we discuss is how a BBH through a weak secular encounter can end up with an eccentricity change $\de$ that
has an opposite sign to what the FO estimate predicts; a feature we denote as `sign reversal'. An example of this is shown in
Fig. \ref{fig:3bodysimJflip}. As seen in the middle plot showing the BBH eccentricity as a function of $t/T_0$, the BBH's $e$ first
increases from $e_0 = 0.99$ to about $e \approx 0.995$, after which the orbit flips (indicated by blue to orange color) and the eccentricity
rapidly decreases and settles at an $\efin \approx 0.9775 < e_0$. The FO estimate completely breaks down in this limit and predicts
a positive $\de$ of  $\approx 0.046$, where the true change is in fact negative, $\de \approx - 0.012$. In this particular example, where the BBH both
undergoes an orbital flip and `exceeds' the $e=1$ limit according to FO theory, the BBH actually ends up with an $\efin < e_{0}$
implying its GW inspiral life time $\tau$ increases as a result of the encounter; something that is far from clear when considering FO theory only. 
In comparison, our SO estimate from Eq. \eqref{eq:de2ord} predicts a change of $\de \approx -0.005$, which correctly is negative, but slightly off as
the initial value of $r_{\rm p}/a_0 = 3.0$ is close to where the secular approximation breaks down. Indeed, we find in this case that the exact
$\de$ depends on the initial angular phase of the BBH, which is a clear indication of a secular breakdown.

How and for what $r_{\rm p}/a_0$ a BBH undergoes a $\de$ sign reversal can e.g. be seen in Fig. \ref{fig:de_etc_rp} in the top plot, where the full (empty) symbols
indicate $\text{sign}(\de) = +1$ ($\text{sign}(\de) = -1$)). The locations of sign reversal are also clearly visible in the middle plot, where the crossing between
a solid line ($\efin$) and its corresponding dashed line ($e_0$) indicates the location where $\de = 0$. For example,
a BBH with an $e_0 = 0.99$ that undergoes a weak encounter with $r_{\rm p}/a_0 \lesssim 3.5$ will end up with a final eccentricity $\efin < e_0$, and vice versa.
As explained, FO theory is not able to resolve any features related to neither the observed `plateau' nor the `$\de$ sign reversal'; however, our SO
theory actually predicts an elegant and simple relation between the location of the plateau, $(r_{\rm p}/a_0)_{\rm plateau}$, and the location
of sign reversal, $(r_{\rm p}/a_0)_{\de = 0}$. As shown in HS19 the two are related by,
\begin{equation}
\frac{(r_{\rm p}/a_0)_{\rm plateau}}{(r_{\rm p}/a_0)_{\de = 0}} = 2^{2/3},
\end{equation}
that is, they are universally connected by the simple factor $2^{2/3} \approx 1.59$.
Finally, if we consider the functional form of $|\de|$ for $(r_{\rm p}/a_0) < (r_{\rm p}/a_0)_{\de = 0}$ we see that it has a different
power law slope than the classical FO prediction $-3/2$. As described in HS19, the slope is different
as this is the region where the SO term from Eq. \eqref{eq:de2ord} dominates the change in eccentricity.
This explains why the slope is observed to be $-3$ as this is exactly the slope associated with $\epsilon^2 \propto (r_{\rm p}/a_0)^{-3}$.

\subsection{Post-Newtonian Effects}\label{sec:Post-Newtonian Effects}

We end this section by discussing the effects from including PN corrects in
the weak interaction itself. PN corrections have been shown to be very important to include for modeling mergers
in the strong interaction few-body problem \citep[e.g.][]{2014ApJ...784...71S, 2018MNRAS.tmp.2223S}, and have likewise been
shown to play a crucial role in the hierarchical Lidov-Kozai problem \citep[e.g.][]{2013ApJ...773..187N}. The question is how important PN corrections are
for our weak interaction scenario. In general, General Relativistic (GR) effects in the PN formalism can 
be divided into two different groups: conservative and dissipative corrections. We discuss both of these
in the following subsections.

\begin{figure}
\centering
\includegraphics[width=\columnwidth]{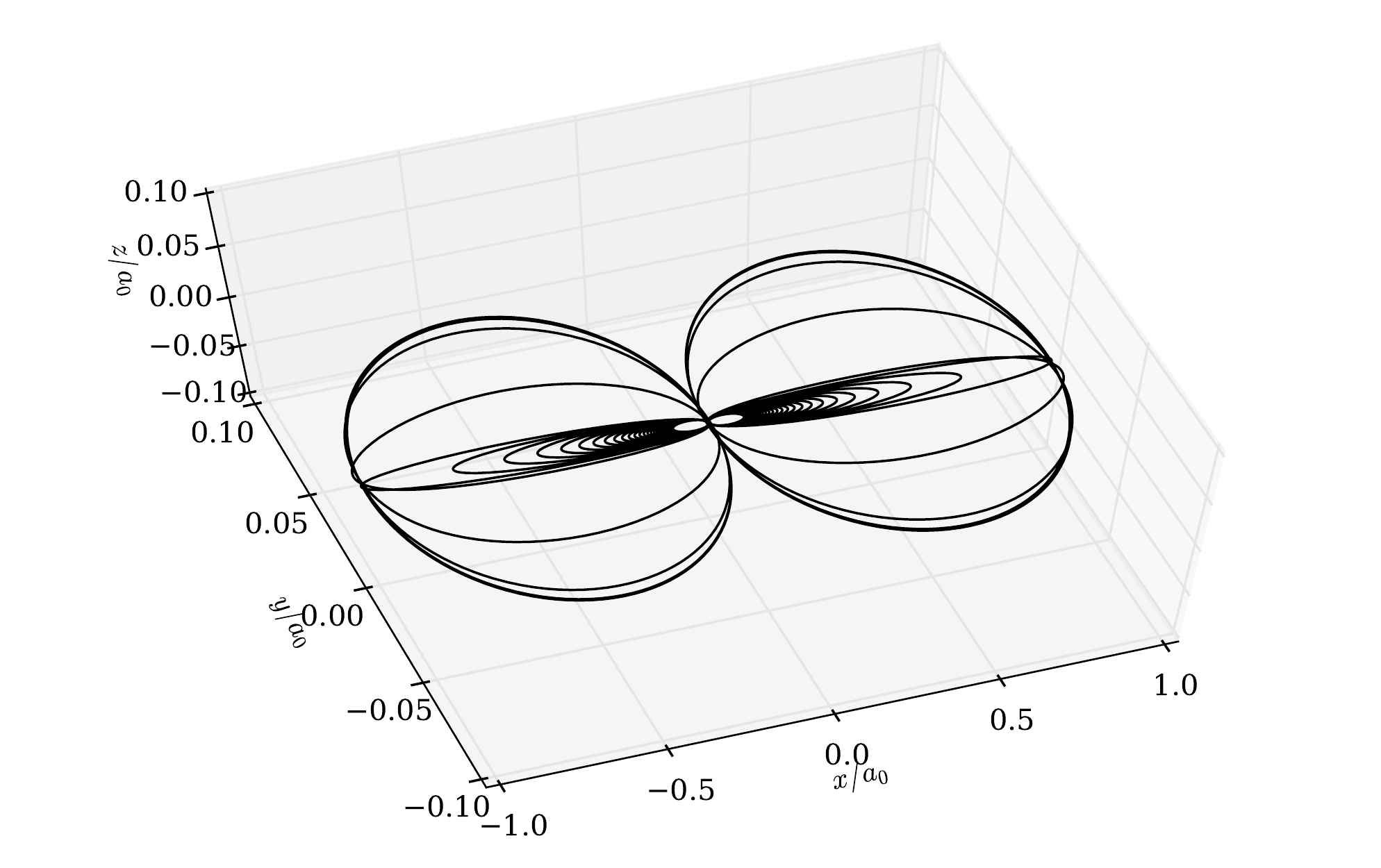}
\caption{Orbital evolution of a BBH undergoing a GW inspiral during a single weak encounter.
The initial conditions are: $r_{\rm p}/a_0 = 5.3$, $a_0 = 0.5$ AU, $e_0 = 0.99$,
and $i = \pi/2$, $\Omega = -\pi/4$, $\omega = 0$. The simulation only includes the 2.5PN term,
as the lower 1,2PN terms lead to precession which make initial conditions ill-defined; however, similar
inspirals form with the 1,2PN terms included.
When comparing to Fig. \ref{fig:de_etc_rp}, one sees that our chosen initial conditions are very close to the
plateau described in Section \ref{sec:Orbital flip and prompt GW Inspiral}. In other words, the weak encounter
leads here to a change $\de \approx 1-e_0$, which leads to a final eccentricity of $\efin \approx 1$.
In this example, the induced GW inspiral time is similar to the time it takes for the encounter to pass the BBH, which explains why we see
the BBH clearly beginning its inspiral during the encounter. This illustrates that prompt GW inspirals
easily can form from distant weak encounters. Because of its short inspiral time, the BBH will appear near the region in-between the LISA and LIGO bands, i.e. near
DECIGO, with notable eccentricity. Further discussions are found in Section \ref{sec:Orbital flip and prompt GW Inspiral} and \ref{sec:Post-Newtonian Effects}.
}
\label{fig:2.5PNinspiralex}
\end{figure}

\subsubsection{Conservative Terms: Precession and Shielding}

The conservative corrections, which are represented by the 1,2 PN terms,
lead to orbital precession. In our case, including precession in the formalism
gives rise to at least two notable effects.
The {\it first effect} is that the initial set of orbital angles (see e.g. Fig. \ref{fig:orbitill}), and especially $\Omega$, are no longer staying constant as the perturber moves
in from infinity due to precession, and a unique outcome is therefore no longer given by only the classical set of ICs.
For a well-defined and unique outcome when PN precession is included, one has to specify exactly where the incoming perturber is
for a given set of orbital angles. This also implies that integrating from time $t \in \pm \infty$ for analytically deriving $\de$ is not possible
as one otherwise does in the Newtonian case (HS19).
For evolving BBHs in clusters this issue of `ill-defined' classical ICs is not a real problem, as the change in angles from PN precession will
be randomized anyway for isotropic random interactions.
The {\it second effect} appears in configurations where the time it takes for the BBH to precess one full $2\pi$ orbit due to PN
apsidal precession \citep[e.g.][]{1972gcpa.book.....W},
\begin{equation}
T_{\rm 1PN} \approx \frac{T_0}{3} \frac{a_0}{\mathscr{R}_{\rm m}}(1-e_0^2),
\label{eq:t1PN}
\end{equation}
where ${\mathscr{R}_{\rm m}}$ denotes the Schwarzschild radius of a BH with mass $m$,
is shorter than the time scale it takes for the encounter to pass the BBH, a time we here refer to as the peri-center passage time $T_{3}$,
\begin{equation}
T_{\rm 3} \approx \sqrt{\frac{r_{\rm p}^{3}}{Gm}},
\label{eq:t3}
\end{equation}
where we have omitted terms of order unity. 
In such cases, the interaction is no longer simply described by an incoming single interacting with a BBH on a closed Keplarian orbit, as the one illustrated in Fig. \ref{fig:orbitill}, but
instead by a single interacting with a `disc' build up by `rosetta' orbits with inner and outer annuli equal to the BBH's peri-center and apo-center, respectively.
This symmetry change from a non-axisymmetric closed orbit to an axisymmetric disc is expected to `shield' the BBH from perturbations, in the same
way as a circular BBH is much harder to perturb than an eccentric \citep[e.g.][]{1996MNRAS.282.1064H}.
Using the above equations \eqref{eq:t1PN} and \eqref{eq:t3}, one finds that such shielding will start to appear for encounters satisfying the following inequality,
\begin{equation}
\begin{aligned}
\frac{r_{\rm p}}{a_0}	& \gtrsim \frac{a_0^{2/3}}{\mathscr{R}_{\rm m}^{2/3}}(1-e_0^2)^{2/3} \\
			   	& \gtrsim 10^{3} \times \left(\frac{a_0}{0.5\text{AU}}\right)^{2/3} \left(\frac{m}{20M_{\odot}}\right)^{-2/3} \left(1-\left({e_0}/{0.99}\right)^{2}\right)^{2/3},
\end{aligned}
\label{eq:rpa1PN}
\end{equation}
where we again have omitted factors of order unity. As seen, even for BBHs with initial high eccentricity, corrections from PN precession are likely
not to play a significant role. However, one might still find resonant couplings between the precession and encounter frequencies
for lower values of $r_{\rm p}/a$, which could lead to PN precession corrections.
We do not have the necessary analytical tools to quantify these highly interesting effects yet, but we will follow up on this in upcoming work.
Also, for this work we updated our PN code to also include the 1,2PN terms, but as also noticed in \citep{2018PhRvD..98l3005R}, it is numerically challenging to
evolve highly relativistic and eccentric BBHs with these corrections. 

\subsubsection{Dissipative Terms: Prompt GW Inspiral}

The dissipative corrections, which first appear at the 2.5PN order, lead to orbital decay for a BBH, which will result in a merger
if its eccentricity is high enough for its GW inspiral life time to be shorter than its dynamical disruption time. As described in Section \ref{sec:Orbital flip and prompt GW Inspiral},
a BBH can be driven to a very high eccentricity $e$ by weak encounters near the plateau, and can in principle reach $e \approx 1$ during orbital flips.
The question is what happens to
such interactions when dissipative corrections, here included by the 2.5 PN term, are included.
An example is shown in Fig. \ref{fig:2.5PNinspiralex}, which shows the orbital evolution of a BBH over the duration of a weak encounter for an interaction
near the plateau. As seen, the weak encounter drives the BBH to such a high eccentricity that the 2.5PN correction
term starts to take over the evolution, which leads to a clear orbital decay during the interaction.
The inspiral time of such binaries are of order the peri-center passage time of the encounter w.r.t. the COM of the BBH, which interestingly
fills in the `gap' in GW peak frequency space between the 3-body mergers and 2-body mergers studied in \citep{2018MNRAS.tmp.2223S}
and later discussed in Section \ref{sec:A Population of Black Hole Binaries}.
Note here that Fig. \ref{fig:2.5PNinspiralex} shows results for a BBH that is similar to the one shown in Fig. \ref{fig:3bodysimJflip},
the only difference is that the one considered in Fig. \ref{fig:2.5PNinspiralex} undergoes a weak encounter from further away; GR effects are therefore not
generally associated with only close strong and weak interactions, but can play a significant role for even very distant perturbers.
Finally, in Fig. \ref{fig:de_etc_rp} results from including the 2.5PN term in weak interactions are shown with {\it black symbols},
where no symbol for the upper two figures, and a cross symbol in the bottom plot,
indicates the BBH merged during the weak interaction. As seen, for interactions where the BBH do not merge during the interaction the final
outcomes are similar to the ones derived using Newtonian dynamics alone. For interactions where the BBH does merge during the interaction,
we find notable differences when including the 2.5PN term in especially the final inspiral time, as seen in the bottom plot. As pointed out in the above paragraph,
such mergers can lead to distinguishable peaks in GW peak frequency space, and are therefore in general important to keep track of.

\subsubsection{Summary}

To conclude, including the 2.5PN dissipative correction is easy and gives rise to potentially important and interesting
prompt GW inspiral mergers forming during or right after the weak interaction.  The conservative 1,2PN corrections are
more difficult to include in the problem, as they also often lead to numerical challenges in the relevant regimes where the 2.5PN correction is also 
important. The 1,2PN terms can lead to shielding effects that protect the binary from changing eccentricity for very distant encounters;
however, using simple scaling arguments, we have proven that this correction is likely not to play a significant role for astrophysical cluster BBHs,
but we will study this in further detail in an upcoming paper. For these reasons, all our studies later on in this paper are based on including the 2.5PN term alone.

\section{Black Hole Binaries in Clusters}\label{sec:Black Holes Binaries in Clusters}

Having addressed in the above Section \ref{sec:A Single Weak Encounter} the outcome of a single weak interaction, we now
turn to the question of what the outcome is of a BBH undergoing a series of weak and strong binary-single interactions inside a dense cluster.

BBHs in dense clusters can be driven to merger during their dynamical hardening through few-body interactions such as binary-single \citep{2014ApJ...784...71S} and
binary-binary \citep{2019ApJ...871...91Z} interactions. Following \citep{2018MNRAS.tmp.2223S}, and later described in Section \ref{sec:Energy Dissipation and Few-body Mergers},
we refer to a BBH that merges during a strong binary-single interaction by a {\it 3-body merger},
and a BBH that merges in-between its strong interactions by a {\it 2-body merger}. As further described in \citep{2018MNRAS.tmp.2223S} and \citep{2019arXiv190102889S},
such mergers can only form if their GW inspiral time
is less or comparable to the characteristic dynamical time of their associated interaction channel, which for 3-body mergers
is $\sim 1$ year (characteristic orbital time of a BBH in a typical GC), and for 2-body mergers is $10^7$ years (characteristic time for a BBH to undergo its
next strong encounter in a typical GC). These characteristic time scales directly maps to how the merging BBHs distribute in
GW peak frequency $f_{\rm GW}$ and orbital eccentricity $e$. Following \citep{2018PhRvD..97j3014S}, the value of $f_{\rm GW}$ for a BBH
with peri-center distance $r_{\rm BH}$ can be approximated by the GW frequency found from assuming the BBH
is circular with a SMA equal to $r_{\rm BH}$. From using that the emitted GW frequency is two times the Keplerian orbital frequency one now finds,
\begin{equation}
f_{\rm GW} \approx \frac{1}{\pi} \sqrt{\frac{2Gm}{r_{\rm BH}^3}}.
\end{equation}
In \citep{2018MNRAS.481.4775D} it was illustrated that a BBH that has to inspiral on a
time scale $\mathcal{T}$, will have an $f_{\rm GW}$ equal to
\begin{equation}
\frac{f_{\rm GW}}{\text{Hz}} \approx 2\cdot10^{-5} \left( \frac{\mathcal{T}}{10^{10}\text{yrs}}\right)^{-3/7} \left( \frac{a}{0.5\text{AU}} \right)^{3/14} \left( \frac{m}{30M_{\odot}}\right)^{-11/14}
\label{eq:fGWTinsp}
\end{equation}
From this relation one concludes that 2-body mergers with $\mathcal{T} \sim 10^{7}$ years will form near the LISA band,
where 3-body mergers with $\mathcal{T}\sim 1$ year form near the LIGO band.
As pointed out in \citep{2018MNRAS.tmp.2223S} and \citep{2018MNRAS.481.4775D}, these dynamically formed BBH
populations will therefore leave unique observational imprints across
the LISA, DECIGO and LIGO bands, which can be used to probe their dynamical history and astrophysical origins. This is highly interesting
and encouraging; however, what has not been discussed in these recent studies is the effect from weak interactions,
and their possible importance for an accurate modeling of the observable $f_{\rm GW}$ and $e$ distributions.
We address this question in this section.

\subsection{Post-Newtonian Cluster Model}

In this section we start by introducing our cluster model, and how we keep track of
different GW merger outcomes. We then describe in simple words the routines of our MC cluster algorithm,
and especially how we include the effects from both GW emission, and strong and weak interactions\footnote{Our developed code
is free to use, and can be obtained by contacting the lead author of the paper.}.
In the last part we present our main results on the role of weak interactions using our introduced MC cluster algorithm.

\subsubsection{Strong Interactions and Dynamical Ejection}\label{sec:Strong Interactions and Dynamical Ejection}

For modeling BBH mergers forming in clusters including weak and strong interactions,
we make use of the following simple model inspired by \citep{2018MNRAS.tmp.2223S}.
We consider a BBH that starts out with an initial SMA $a$ that is slightly below the hard-binary (HB) value given by
\begin{equation}
a_{\rm HB} \approx \frac{3}{2}\frac{Gm}{v_{\rm dis}^{2}},
\end{equation}
We imagine this BBH lives in a population of single BHs, all with the same mass $m$,
characterized by a constant number density $n_{\rm s}$, velocity dispersion $v_{\rm dis}$, and escape velocity $v_{\rm esc}$.
The BBH will undergo both strong and weak interactions defined in the following way:
We denote an interaction a {\it strong interaction} if the incoming single BH passes
the BBH on an orbit with peri-center distance $r_{\rm p} < \mathscr{C}a$, where our fiducial value of $\mathscr{C}$ is here $\mathscr{C}=2$, and
$a$ is the SMA of the BBH at the time the single passes the BBH. An interaction with $r_{\rm p}>\mathscr{C}a$ will instead be labeled a {\it weak interaction}.
We assume that each strong interaction leads to a constant decrease in the
SMA of the BBH from $a$ to $a_{\rm enc}$ as, 
\begin{equation}
a_{\rm enc} = a \times \delta,
\end{equation}
where we use $\delta = 7/9$, which equals the average value assuming the chaotic binary-single interaction can be modeled by
an ergodic distribution as discussed in \citep{2018PhRvD..97j3014S}.
At the same time we assign the BBH a new eccentricity $e$ that is drawn from a so-called thermal distribution $P(e) = 2e$ \citep{Heggie:1975uy}.
We choose this simple model for strong interactions due to the following three reasons.
$(i)$ It provides full analytical solutions to both the stationary BBH population \citep[e.g.][]{2019PhRvD..99f3006S} and the highly eccentric evolving population across
all GW bands \citep[e.g.][]{2018PhRvD..97j3014S, 2019arXiv190102889S}. This allows us to not only cross check all our algorithms,
but also to extract small changes coming from additional terms such as weak interactions.
$(ii)$ The simple $\delta$-model has been used in all our previous studies, which allows for an easy comparison. Despite its simplicity,
our previous results \citep[e.g.][]{2018MNRAS.tmp.2223S} are in surprisingly good agreement with more sophisticated MC techniques \citep[e.g.][]{2019PhRvD..99f3003K}.
$(iii)$ Adding weak interactions into the evolution of BBHs requires in principle large computational resources, as weak interactions need to be performed $\gg 1$ times
in-between each strong interaction. Therefore, partly due to computational limitations, we restrict ourself to our simple $\delta$-model
combined with additional hybrid techniques, as described later.

This process of a gradual decrease of the BBH's SMA is loosely referred to as {\it hardening}. In the point-particle Newtonian limit this will always end with
a dynamical ejection of the BBH when its SMA reaches the following critical value,
\begin{equation}
a_{\rm ej} \approx \frac{1}{6} \left(\frac{1}{\delta} - 1\right) \frac{Gm}{v_{\rm esc}^2},
\end{equation}
which is where the recoil velocity from the 3-body interaction equals the escape velocity of the cluster \citep{2018PhRvD..97j3014S}.
This classical hardening process therefore leads to an ejection of BBHs with a SMA $\sim a_{\rm ej}$
and an eccentricity distribution that is thermally distributed. As shown by \citep{2000ApJ...528L..17P, 2016PhRvD..93h4029R}, this population of
ejected BBHs can explain at least part of the observed LIGO merger rate, but as pointed out by e.g. \citep{2018PhRvD..97j3014S, 2018MNRAS.tmp.2223S}, this picture
is highly incomplete as it only includes (energy conserving) Newtonian dynamics. This is described in greater detail below.

\subsubsection{Energy Dissipation and Few-body Mergers}\label{sec:Energy Dissipation and Few-body Mergers}

When GR dissipative terms -- where the leading term is the 2.5PN term in the PN expansion formalism -- 
are included in the dynamics, a BBH can also merge inside its cluster either {\it during} or {\it in-between} its strong interactions.
We describe our methods for modeling these outcomes in the following paragraphs.

{\it 3-body Mergers}: For determining if a BBH merges during a strong 3-body interaction,
we model this often highly chaotic and resonating state using the approach put forward in \citep{2018PhRvD..97j3014S}.
The interaction here is divided up into $N_{\rm IMS}$ intermediate states (IMSs), each of which is described by a BBH with a bound single BH.
For each IMS we assign the corresponding BBH an eccentricity sampled from the thermal distribution $P(e) =2e$, but keep
the SMA fixed to its initial value $a_{0}$. To determine if the BBH undergoes a GW inspiral merger during this IMS, i.e. merge while the
third objects is still bound to it, we compare the energy radiated over one orbit of the BBH,
\begin{equation}
\Delta{E}_{\rm GW} \approx (85\pi/12)G^{7/2}c^{-5}m^{9/2}r_{\rm BH}^{-7/2},
\end{equation}
to the total orbital energy of the bound 3-body state,
\begin{equation}
E_{\rm B} \approx Gm^{2}/(2a_{0}).
\end{equation}
If $\Delta{E}_{\rm GW} > E_{\rm B}$ we label the IMS assembled BBH as a 3-body merger. This energy threshold is equivalent of saying that
the BBH will undergo a GW inspiral merger during the interaction if its peri-center distance $r_{\rm BH} < r_{\rm 3cap}$, where $r_{\rm 3cap}$ is a
`3-body capture distance' given by \citep{2018PhRvD..97j3014S},
\begin{equation}
r_{\rm 3cap} \approx {\mathscr{R}_{\rm m}} \times \left({a_0}/{{\mathscr{R}_{\rm m}}}\right)^{2/7}.
\end{equation}
If instead $r_{\rm BH} > r_{\rm 3cap}$ the BBH does not merge during the considered IMS.
We repeat this process, i.e. first assign the IMS assembled BBH an eccentricity from $P(e) = 2e$ and then compare its $r_{\rm BH}$ to $r_{\rm 3cap}$,
up to $N_{\rm IMS} =20$ times per interaction, which is the average number of IMSs assembled in a chaotic equal mass 3-body interaction \citep{2018PhRvD..97j3014S}.
The total probability for a BBH to merge during a 3-body interaction, $P_{\rm 3m}$, i.e. to be
formed with orbital parameters such that $r_{\rm BH} < r_{\rm 3cap}$, can in this model be expressed analytically as follows \citep{2018PhRvD..97j3014S}, 
\begin{equation}
P_{\rm 3m} \approx \frac{2r_{\rm 3cap}}{a_0} \times N_{\rm IMS}.
\end{equation}
As illustrated in \citep{2018PhRvD..97j3014S}, by integrating this probability up over a BBHs hardening binary-single interactions from $a_{\rm HB}$ to $a_{\rm ej}$,
one finds that the total probability for a BBH to merge during a 3-body interaction relative to merging outside the cluster within a
Hubble time is $\approx 10\%$.

{\it 2-body Mergers}: For keeping track of the formation of 2-body mergers, we evolve the orbital parameters $a,e$ of the BBH
in question in-between each interaction according to the equations given by \citep{Peters:1964bc}.
If at any given time the GW inspiral life time of the BBH right after it has finished an interaction, $\tau$,
is less than the time it takes before it undergoes its next interaction, $t_{\rm int}$, we stop the code
and label the outcome a 2-body merger. To calculate the time before the next interaction, we use the same approach as
for Eq. \eqref{eq:tSE}. Assuming that we keep track of single encounters up to a maximum peri-center distance of $R_{\rm p}$,
the time in-between interactions can be approximated by,
\begin{equation}
t_{\rm int} \approx  \frac{1}{6 \pi G} \frac{v_{\rm dis}}{n_{\rm s} m R_{\rm p}}.
\label{eq:tint}
\end{equation}
As argued in Section \ref{sec:Distant weak encounters}, one does expect the results to converge for $R_{\rm p}$ going to infinity.
We will address this issue later in Section \ref{sec:A Population of Black Hole Binaries}.

\subsubsection{BBH Evolution with Weak Interactions}\label{sec:BBH Evolution with Weak Interactions}

We now describe our numerical scheme used in this paper for modeling the evolution of BBHs inside a dense stellar cluster
with the inclusion of weak interactions. It expands on a similar model put forward by \citep{2018PhRvD..97j3014S, 2018MNRAS.tmp.2223S, 2019arXiv190102889S}.

\begin{figure}
\centering
\includegraphics[width=\columnwidth]{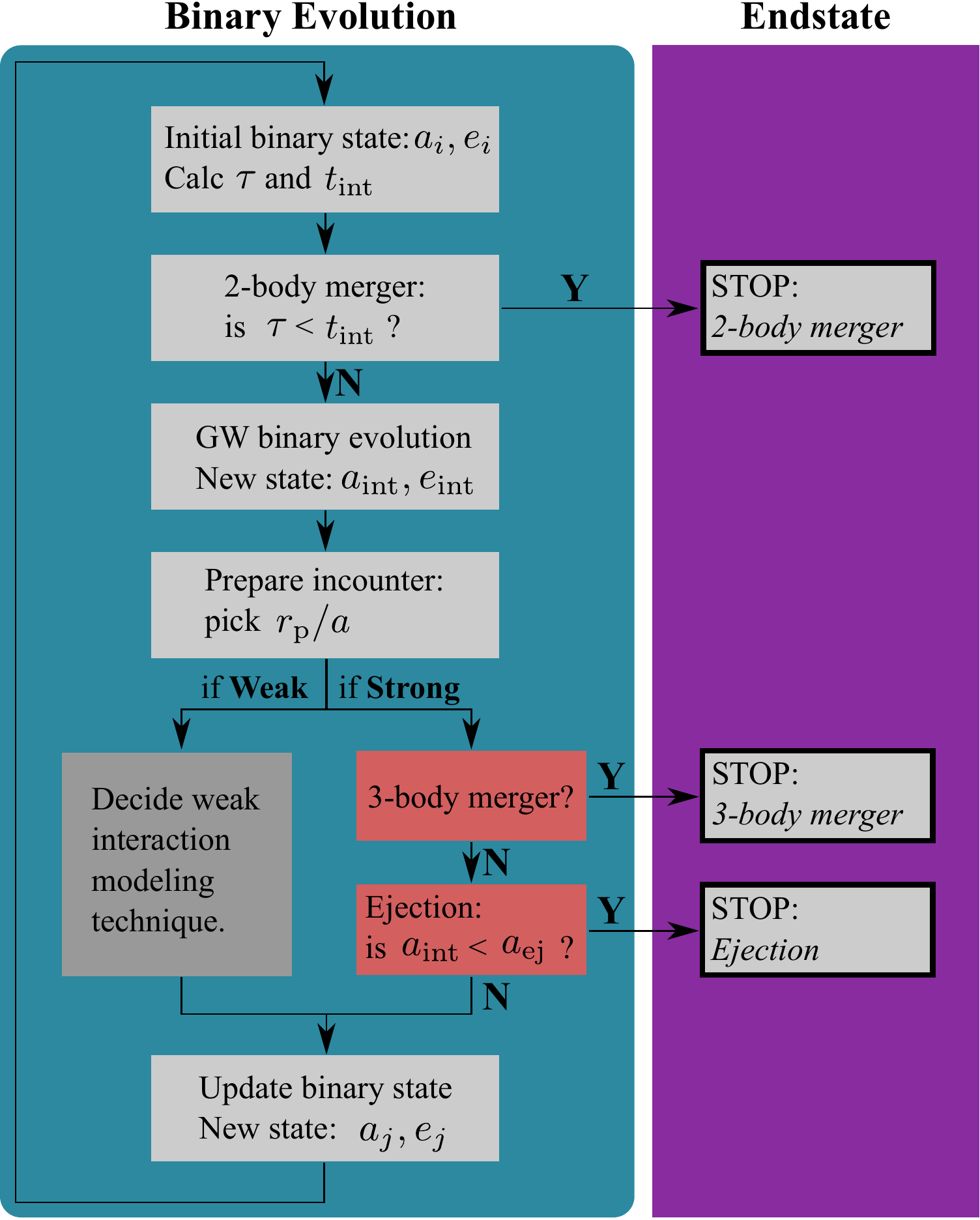}
\caption{Flowchart illustrating the main steps and logic of our BBH evolution algorithm described in Section \ref{sec:BBH Evolution with Weak Interactions}.
The left {\it blue box} evolves the BBH from state $i$ to state $j$, which includes GW and dynamical evolution. Based on what value that is picked for the peri-center
distance of the incoming single w.r.t. the BBH COM, $r_{\rm p}/a$, the interaction is described either by a weak (if {\bf Weak}) or a strong interaction (if {\bf Strong}).
The righ {\it purple box} highlights the three main endstates we keep track of: 2-body merger, 3-body merger, and dynamical ejection.
The flowchart only shows the most important steps of our routine, and includes therefore not additional stopping criteria such as BBH merger forming during a
weak interaction; however, our full procedure is described in detail in Section \ref{sec:BBH Evolution with Weak Interactions}.}
\label{fig:flowchart}
\end{figure}

We start by first choosing the maximum relative peri-center distance within which we model the effect from interactions, $R_{\rm p}/a$, where
$a$ is the SMA of the BBH in question. This rescaled value stays constant, but the physical value of $R_{\rm p}$ will decrease as the BBH hardens inside
its cluster. If only strong interactions are included, then $R_{\rm p}/a \sim 1$. Therefore, a value of $R_{\rm p}/a \gg 1$ will
allow the orbital parameters of the BBH to be affected by a much wider and numerous range of interactions compared to
the previously studied strong interaction case. With this choice of $R_{\rm p}/a$, we follow the evolution of a BBH from its initial SMA, $a_{\rm in} \sim a_{\rm HB}$,
towards its final value, $a_{\rm fin} \sim a_{\rm ej}$, assuming all properties of the cluster stay constant. As $R_{\rm p}/a$ takes a finite
value we are able to evolve the BBH in finite steps. In the following we describe how we evolve the BBH from step $i$ to its next step $j$.
See also Fig. \ref{fig:flowchart} which shows a simplified flowchart of our algorithm.
\\
\\
{{\bf (1) Check for 2-body merger}: Based on the BBH's orbital values at step $i$, referred to as $\{a_i, e_i\}$, and the value of $R_{\rm p}/a$, we first derive
$\tau$ and $t_{\rm int}$ using Eq. \eqref{eq:tGW} and \eqref{eq:tint}, respectively. If $\tau < t_{\rm int}$ we stop the
code and record the BBH as a 2-body merger. If instead $\tau > t_{\rm int}$, we proceed as follows.}
\\
\\
{{\bf (2) GW orbital evolution}: If no 2-body merger has occurred, we evolve the BBH's orbital parameters
using the equations from \citep{Peters:1964bc} over the time interval it takes for it to undergo its next encounter $t_{\rm int}$.
During this evolution the orbital parameters of the BBH evolve from $\{a_i, e_i\}$ to $\{a_{\rm int}, e_{\rm int}\}$.}
\\
\\
{{\bf (3) Pick encounter distance}: We now prepare for the encounter. For this we start by randomly picking a peri-center distance
for the encounter w.r.t. the COM of BBH, $r_{\rm p}/a_{\rm int}$, in the range $[0: R_{\rm p}/a_{\rm int}]$
assuming the encounters are isotropically distributed at infinity (see e.g. \citep{2014ApJ...784...71S}).
At this step we just record the value $r_{\rm p}/a_{\rm int}$.}
\\
\\
{{\bf (4) Classify and perform encounter}: At this step we start by classifying the encounter type from the above step. We do this as follows.
\\
{\it - Strong Interaction}: If $r_{\rm p}/a_{\rm int} < \mathscr{C}$, we model the interaction as a strong interaction. 
To evaluate the corresponding outcome we first estimate if the interaction results in a $3$-body
merger using the IMS-decomposition approach described in Section \ref{sec:Energy Dissipation and Few-body Mergers}.
If a $3$-body merger does not form, we model the interaction
outcome using our simple $\delta$-model described in Section \ref{sec:Strong Interactions and Dynamical Ejection}.
If at this point $a_{\rm int} < a_{\rm ej}$, we stop the code after the interaction and record the BBH as being ejected from the cluster.
\\
{\it - Weak Interaction}: If instead $r_{\rm p}/a_{\rm int} > \mathscr{C}$, we model the interaction as a weak interaction.
For this we start by picking a random configuration for the angles $\{i, \omega, \Omega\}$ assuming an isotropic distribution.
We then derive the outcome using one of the following ways.
$(i)$: Use our few-body code described in \citep{2017ApJ...846...36S}.
$(ii)$: Use our SO result from Eq. \eqref{eq:de2ord}.
$(iii)$: Use a `hybrid approach' where we combine $(i)$ and $(ii)$.
As discussed later, option $(i)$ is numerically challenging, but straight forward. Option $(ii)$ is in-accurate for $r_{\rm p}/a_{\rm int} \sim \mathscr{C}$,
but extremely fast. Therefore, option $(iii)$ is often the best option, but it requires bridging $(i)$ and $(ii)$.}
\\
{\it - Final parameters}: If the BBH has not merged at this point its final set of orbital values will be
$\{a_{j}, e_{j}\} = \{a_{\rm int} + \Delta{a}, e_{\rm int} + \Delta{e}\}$, with $\Delta{a}$ and $\Delta{e}$
derived as just described.
\\
\\
{{\bf (5) Next step and termination}: For a BBH that has survived all interactions and is still inside the cluster,
we end this `one interaction'-routine from step $i$ to $j$ by simply putting $\{a_{\rm j}, e_{\rm j}\} \rightarrow \{a_{\rm i}, e_{\rm i}\}$.
After this we go back to the start and repeat the process. We keep doing this until the BBH either merges inside the cluster
or is ejected.} 
\\
\\
Using this algorithm we can quickly build up a distribution of BBH properties showing e.g. the distribution of BBH eccentricity
$e$ across different $f_{\rm GW}$ bands from LISA to LIGO. Our model does represent a simplified picture of how BBHs undergo dynamical interactions
in dense stellar systems; however, our extremely fast approach has shown to give surprisingly accurate predictions
for how BBHs merge and distribute across both $f_{\rm GW}$ and $e$. Our model can easily be improved,
but its simplicity as well as its corresponding list of available analytical solutions makes it an ideal testbed for including new effects,
as the ones we consider in this paper. Below we study outcomes from this model.

\subsection{Results from Including Weak Encounters}\label{sec:Results from Including Weak Encounters}

In this section we present our main results for how BBHs interact and merge in clusters
under the influence of both weak and strong interactions. After defining our fiducial parameters
and approach below, we study the dynamical history of a single BBH. After this we
move to results derived from a large sample of evolved BBHs using MC techniques, from which we derive
$f_{\rm GW}$ and $e$ distributions relevant for especially LISA. As illustrated, we find clear observational
differences when weak interactions are included. 

\subsubsection{Fiducial Parameters}\label{sec:Fiducial parameters and model}

In all our models we use the algorithm described in the above
Section \ref{sec:BBH Evolution with Weak Interactions}. If not stated otherwise, we
further assume the following set of `fiducial model' parameters:
$m=20M_{\odot}$, $n_{\rm s} = 10^{5}\ \text{pc}^{-3}$, $v_{\rm dis} = 10\ \text{kms}^{-1}$,
$v_{\rm esc} = 50\ \text{kms}^{-1}$, $R_{\rm p}/a = 50$, and $\mathscr{C}  = 2$.
These are chosen for easy comparison to previous work.
For our main results presented below, we also employ the following `hybrid approach'
for accurate and fast modeling of the outcome from weak encounters: 
for encounters with $2 < r_{\rm p}/a < \max(5, (r_{\rm p}/a_0)^{\rm BR}_{\rm FO})$ we use a full 2.5PN three-body integration to evaluate the outcome, where for
$r_{\rm p}/a > \max(5, (r_{\rm p}/a_0)^{\rm BR}_{\rm FO})$ we use HS19 given by Eq. \eqref{eq:de2ord}.
In Section \ref{sec:Full Analytical Approach} we compare results from this hybrid approach to results using HS19 only.

\subsubsection{A Single Black Hole Binary}\label{sec:A Single Black Hole Binary}

\begin{figure*}
\centering
\includegraphics[width=0.8\textwidth]{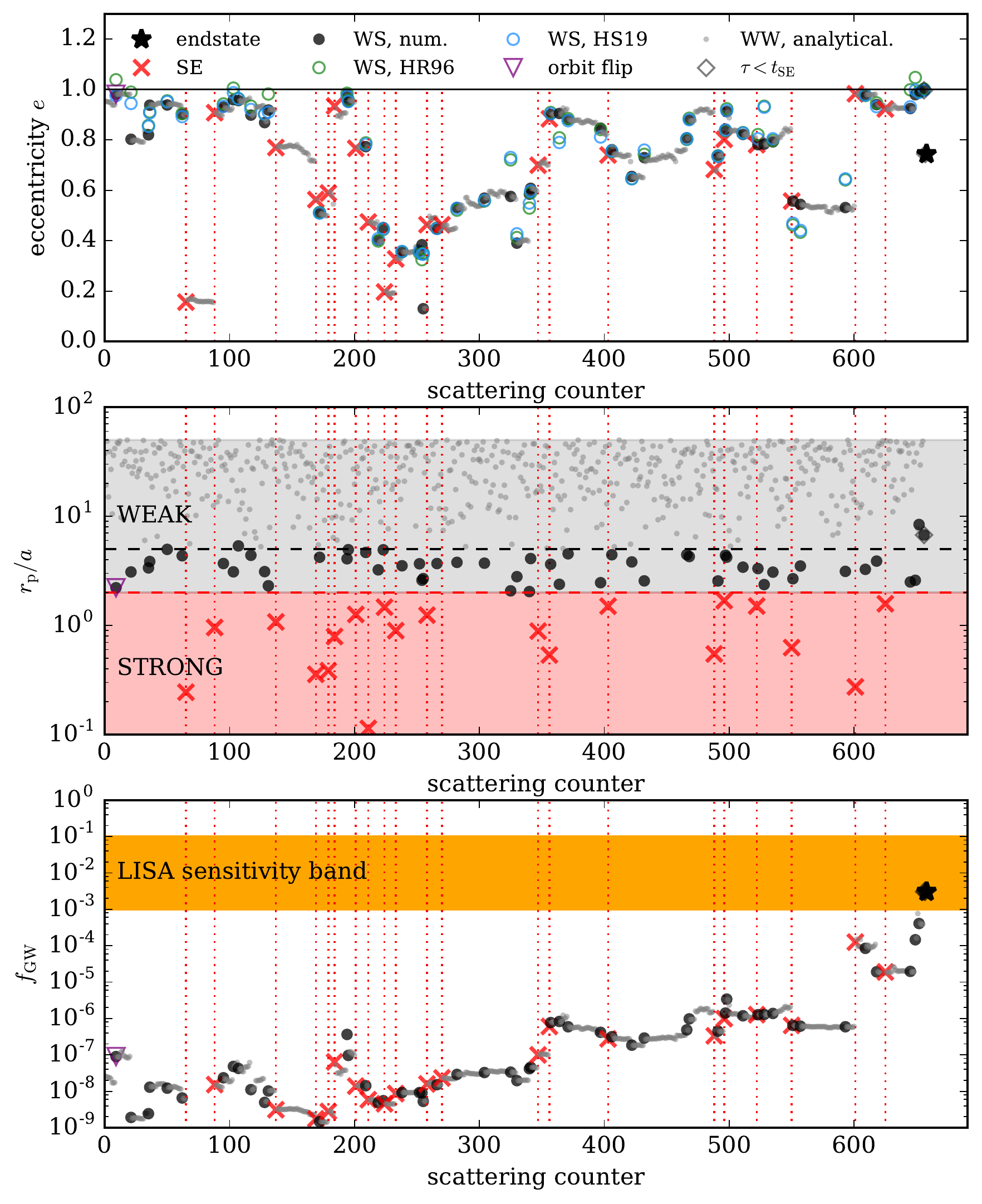}
\caption{Dynamical history of a BBH evolving inside a cluster through equal mass weak and strong three-body interactions.
For this simulation we have used our numerical scheme outlined in Section \ref{sec:BBH Evolution with Weak Interactions}, with the inclusion of the following
hybrid approach as further described in Section \ref{sec:A Single Black Hole Binary}. In short: 
For {\it strong encounters} (SE), characterized here by $r_{\rm p}/a < 2$,
we evolve the BBH assuming the SMA $a \rightarrow \delta a$, and $P(e) = 2e$.
For {\it weak-strong encounters} (WE), characterized here by $2 < r_{\rm p}/a < 5$,
we use our 2.5PN code to evaluate the outcome.
For {\it weak-weak encounters} (WW), characterized here by $r_{\rm p}/a > 5$, we
assume $\Delta{a} = 0$ and use Eq. \eqref{eq:de2ord} to derive $\de$.
For the whole evolution we use the fiducial cluster values listed in Section \ref{sec:Results from Including Weak Encounters}.
The BBH evolves from its initial values from left to right as a function of scatterings (scattering counter), and ends by merging inside
the cluster in-between its interactions (2-body merger). Each point in the plots shows the resulting outcome from the corresponding scattering.
We use the following labeling to highlight different encounters and outcomes:
{\it Red cross}: Strong-encounter (SE).
{\it Black dot}:  Weak-strong encounter (WS); result from our numerical integration.
{\it Green circle}:  Weak-strong encounter (WS); result from HR96.
{\it Blue circle}:  Weak-strong encounter (WS); result from HS19.
{\it Purple triangle}: BBH orbital flip.
{\it Grey dot}:  Weak-weak encounter (WW).
{\it Grey diamond}: The BBH's inspiral time, $\tau$ is $<$ than the time it takes for the next strong encounter to happen, $t_{\rm SE}$.
{\it Black star}: Endstate; at this point the BBH will inspiral in near isolation with no further perturbations ($\tau < t_{\rm int}$).
Each of the {\it red dotted} vertical lines indicate a SE; evolution in-between these lines is therefore due to weak encounters and
GW emission only.
The {\it top plot} shows the BBH's eccentricity $e$. The {\it middle plot} shows $r_{\rm p}/a$ for each scattering, with weak and strong
scattering zones highlighted (grey/red). The {\it bottom plot} shows $f_{\rm GW}$ with the LISA band overplotted.
Results are discussed in Section \ref{sec:A Single Black Hole Binary}.
}
\label{fig:fGWdist}
\end{figure*}

We start by studying the evolution of a single BBH undergoing both weak and strong encounters inside a dense cluster. Results are presented in
Fig. \ref{fig:fGWdist}, which shows in the top, middle, and bottom plots the BBH's eccentricity $e$, the value of $r_{\rm p}/a_0$,
and the BBH's peak GW frequency $f_{\rm GW}$, respectively, as a function of the number of scatterings, $N_{\rm sc}$ (scattering counter). 
For this modeling we have used our numerical scheme described in Section \ref{sec:BBH Evolution with Weak Interactions},
with the hybrid approach and fiducial cluster parameters listed in the above Section \ref{sec:Fiducial parameters and model}.

As seen, the BBH occasionally undergoes notable evolution in eccentricity in-between its strong interactions (red crosses).
For example, the BBH evolves from an eccentricity of $\sim 0.8$ to $\sim 0.9$ over the range $N_{\rm sc} \approx 400-500$ entirely through weak interactions (grey dots).
A relevant question is here if weak interactions therefore are expected to lead to an observed eccentricity distribution that is notably different from the thermal
distribution that often follows from strong scatterings. However, by studying a large number of runs of BBHs evolving through
both weak and strong interactions, we did not find any significant differences when weak interactions are included.
We will however study this in more detail in a upcoming paper. See also \citep{2019ApJ...872..165G} for a similar analysis.

The importance of including weak encounters becomes clear when considering the end of the evolution of the BBH shown in Fig. \ref{fig:fGWdist}.
As seen, the last strong interaction leaves the BBH with an eccentricity of $e \approx 0.9$, which is not even high enough for it to merge before its next
strong interaction (if the BBH has an inspiral time $\tau < t_{\rm SE}$, where $t_{\rm SE}$ is the time interval before the next strong encounter (see Eq. \eqref{eq:tSE}),
the state is marked with a grey diamond). However, the BBH is instead driven to merge entirely
through weak interactions, each of which changes the BBH's eccentricity until $\tau < t_{\rm int}$. This evolution also brings the BBH
into the LISA band, as seen in the bottom plot, which will make it appear as a high eccentricity LISA source. Such eccentric sources are not a natural outcome
of isolated BBH evolution, and therefore provide a test of the cluster channel \citep{2018MNRAS.tmp.2223S, 2019PhRvD..99f3003K}.

Finally, the top plot shows how the analytical estimates HR96 (green circles), and HS19 (blue circles), compare to our
full numerical results (black dots). As seen, both HR96 and HS19 provide excellent estimates when $r_{\rm p}/a \gg 1$; however, both start to break down, as expected,
when approaching the strong interaction limit. The important difference between HR96 and HS19 is seen for BBHs that
scatter around the $e \approx 1$ limit. For example, HR96 clearly gives rise to several unphysical
outcomes with $e>1$ when the BBH is already highly eccentric; a problem that is resolved by using HS19 due to its corrections from the SO term.
The high eccentricity region is the most interesting and relevant for understanding and modeling how BBHs are driven to merge, which again
illustrates the importance of using corrections similar to the one encoded in HS19.

\subsubsection{A Population of Black Hole Binaries}\label{sec:A Population of Black Hole Binaries}

We now turn to the study of what GW signals to expect from a population of BBHs evolving through weak and strong interactions, similar to the population that we
potentially observe from GCs throughout the nearby observable universe. For this, we here model the dynamics of an ensemble
of $15000$ individual BBHs, and follow their evolution in a similar way as the one described in the above Section \ref{sec:A Single Black Hole Binary}.
We record 2-body and 3-body BBH mergers, as well as properties for the ejected BBH population.
From this data we derive $f_{\rm GW}$ and $e$ distributions, as described in the following
sections. In this sub-section all our results are based on the fiducial values and
hybrid approach outlined in Section \ref{sec:Fiducial parameters and model}.

\begin{figure}
\centering
\includegraphics[width=\columnwidth]{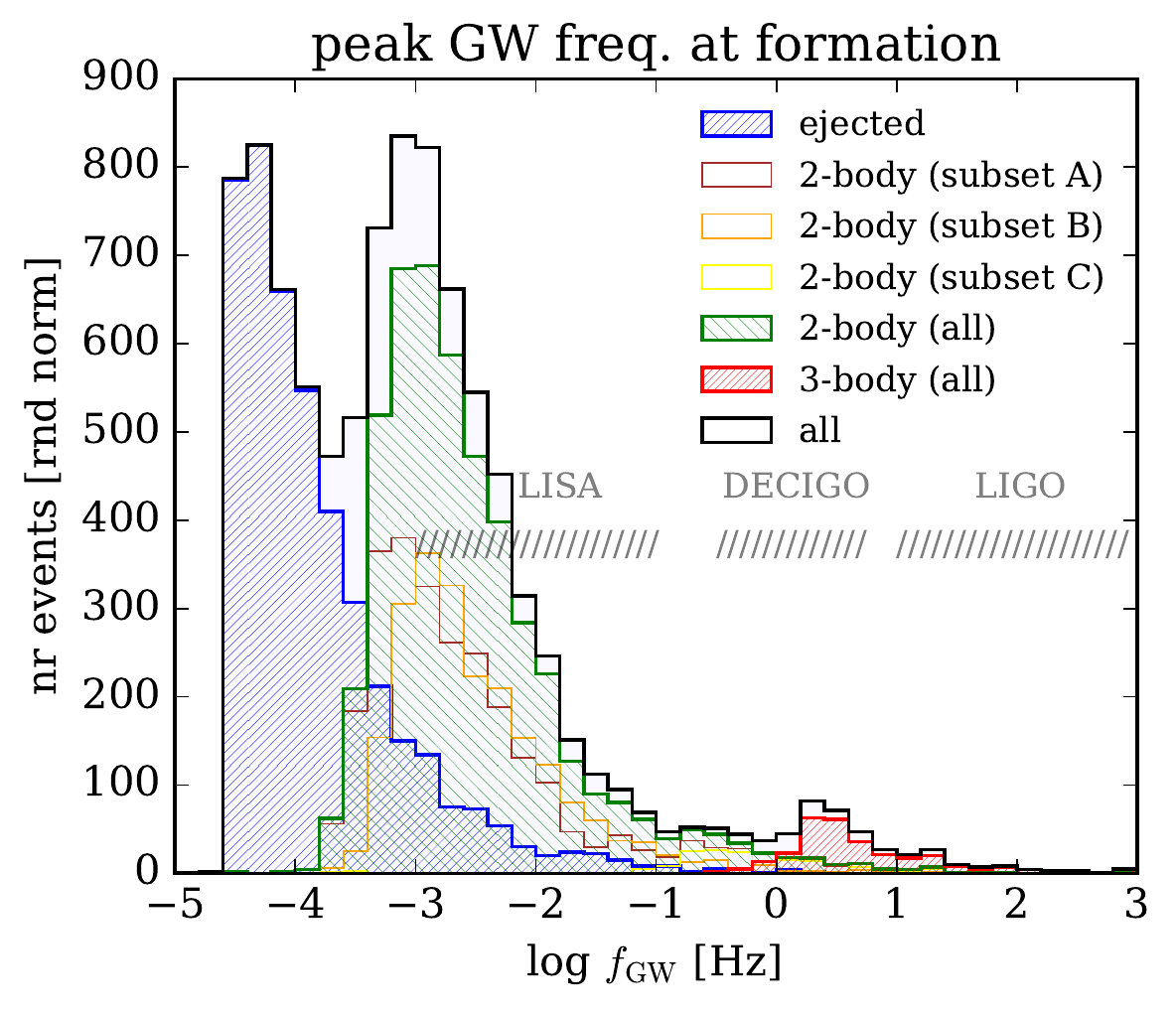}
\includegraphics[width=\columnwidth]{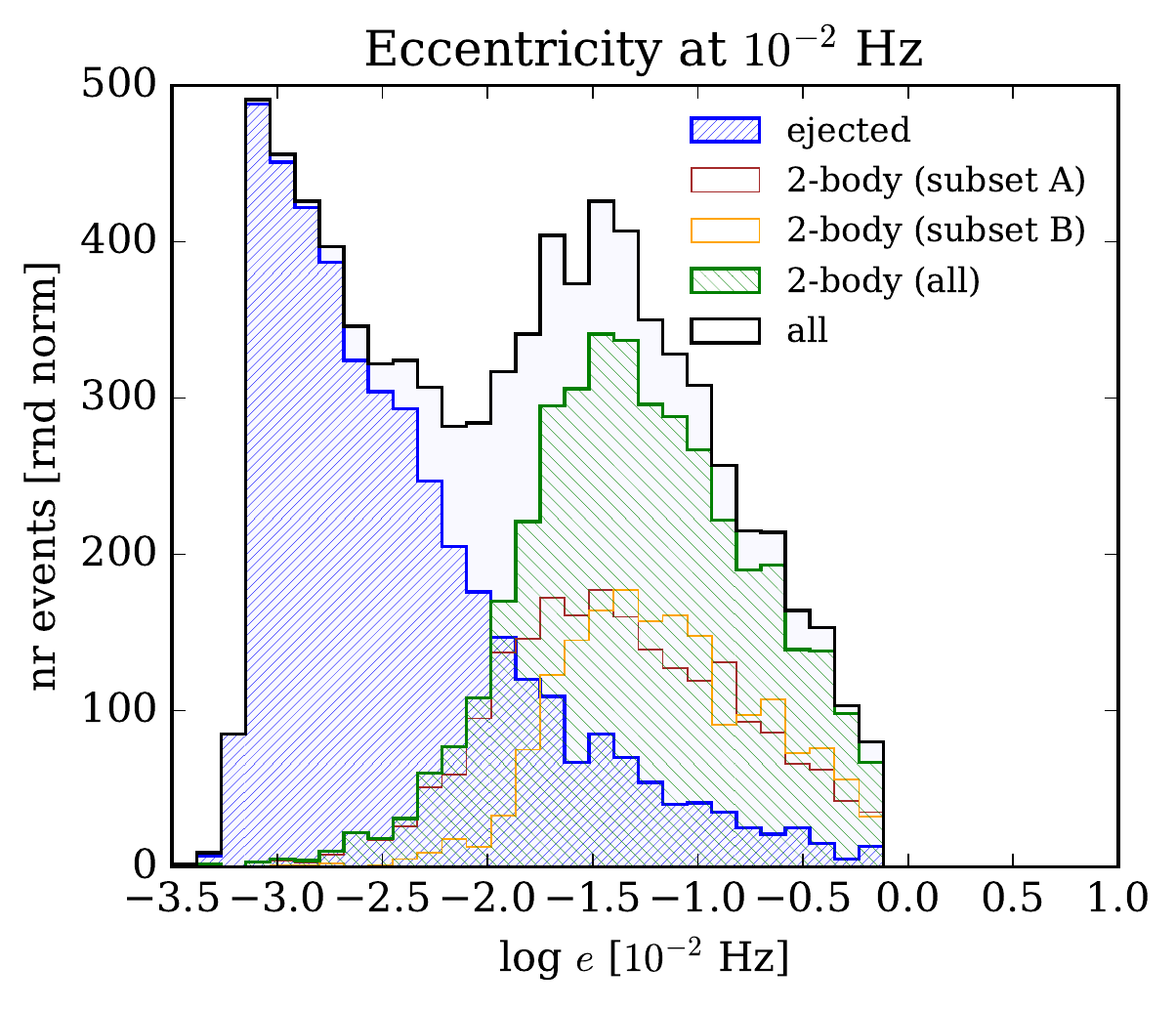}
\caption{Distribution of GW peak frequency $f_{\rm GW}$ ({\it top plot}), and
BBH orbital eccentricity evaluated at $f_{\rm GW} = 10^{-2}$ Hz ({\it bottom plot}), derived
from $15000$ individual runs of BBHs undergoing weak and strong scatterings
inside a dense cluster. For this, we used the numerical scheme outlined in
Section \ref{sec:BBH Evolution with Weak Interactions}, using our `fiducial cluster parameters' and
`hybrid approach' described in Section \ref{sec:Results from Including Weak Encounters}.
The distributions are divided into the following sub-categories:
{\it Ejected}: BBHs ejected from the cluster with an inspiral time $\tau < t_{\rm H}$.
{\it 2-body (Subset A)}: 2-body mergers where the BBH from the last recorded strong-encounter
has $\tau > t_{\rm SE}$, i.e. the BBH merges due to weak encounters. 
{\it 2-body (Subset B)}: 2-body mergers where the BBH from the last recorded strong-encounter
has $\tau < t_{\rm SE}$. 
{\it 2-body (Subset C)}: BBHs that merge during a weak interaction.
{\it 2-body (all)}: All 2-body mergers, i.e. subset $A + B + C$.
{\it 3-body (all)}: All 3-body mergers.
{\it All}: All mergers, i.e. ejected + 2-body + 3-body.
The distribution of $f_{\rm GW}$ is the value at `formation', i.e. right when we identify the BBH in question
to either have been ejected, has an inspiral time $\tau < t_{\rm int}$ for 2-body mergers, or a peri-center distance
during a 3-body interaction $r_{\rm BH} < r_{\rm 3cap}$ for 3-body mergers.
Results are discussed in Section \ref{sec:A Population of Black Hole Binaries}.
}
\label{fig:fGWdistloge102}
\end{figure}

We first consider Fig. \ref{fig:fGWdistloge102}, which shows the distribution of $f_{\rm GW}$ (top plot) and the eccentricity distribution
at $10^{-2}$ Hz (bottom plot) of the merging BBH population. As described in the caption, $f_{\rm GW}$ is the GW peak frequency recorded at the
time of `formation', i.e. right when either the BBH in question is ejected from the cluster for
ejected mergers, $\tau < t_{\rm int}$ for 2-body mergers, or right when $r_{\rm BH} < r_{\rm 3cap}$ for 3-body mergers.
This distribution is therefore not universal, as $t_{\rm int}$ depends on $R_{\rm p}$ (a higher $R_{\rm p}$
will allow each BBH to inspiral over a longer time scale before the code halts, which will drive the 2-body merger distribution slightly to the right). However, it
only weakly depends on $R_{\rm p}$ as each BBH inspirals with near constant $f_{\rm GW}$, and the plot therefore still provides
some useful insight into how the different outcomes distribute and differ from each other. As seen, ejected mergers tend to form below both the LISA and
the LIGO bands, and will therefore almost circularize before being observable, whereas 2-body mergers will appear eccentric in LISA, and 3-body mergers in LIGO.
As described in \citep{2018MNRAS.481.4775D}, and illustrated by Eq. \eqref{eq:fGWTinsp}, this splitting of outcomes
across $f_{\rm GW}$ is linked to the characteristic inspiral times of each population.
The bottom plot of $e$ at $10^{-2}$ Hz is in principle not dependent on $R_{\rm p}$, as long as the problem converges
with increasing $R_{\rm p}$. From this plot its clear that the 2-body mergers dominate the population of eccentric sources in the LISA band.
The different colored lines in both plots illustrate different dynamical subsets, as described in the figure caption.
Starting with comparing {\it Subset A} (last strong interaction has $\tau > t_{\rm SE}$, i.e. BBHs are here driven to merger through one or more weak encounters)
and {\it Subset B} (last strong interaction has $\tau < t_{\rm SE}$, i.e. the BBHs will here still be labeled mergers using a strong interactions code only),
we see that there is no major difference between the two. It will therefore generally be hard to tell the difference between strong and weak interaction
driven BBH mergers from a few LISA observations alone (in this statement we have ignored rare exotic mergers such as a BBH promptly driven to merger 
through an interplay between PN corrections and a dynamical flip-orbit during a weak interaction). The only major difference across $f_{\rm GW}$
from the weak interaction channel is the appearance of {\it Subset C}, which originates from BBHs that merge during
the weak encounter (see e.g. Fig. \ref{fig:2.5PNinspiralex}). As seen, these mergers appear at relatively high frequencies where they make up their
own `bump' slightly below the 3-body mergers.

\begin{figure}
\centering
\includegraphics[width=\columnwidth]{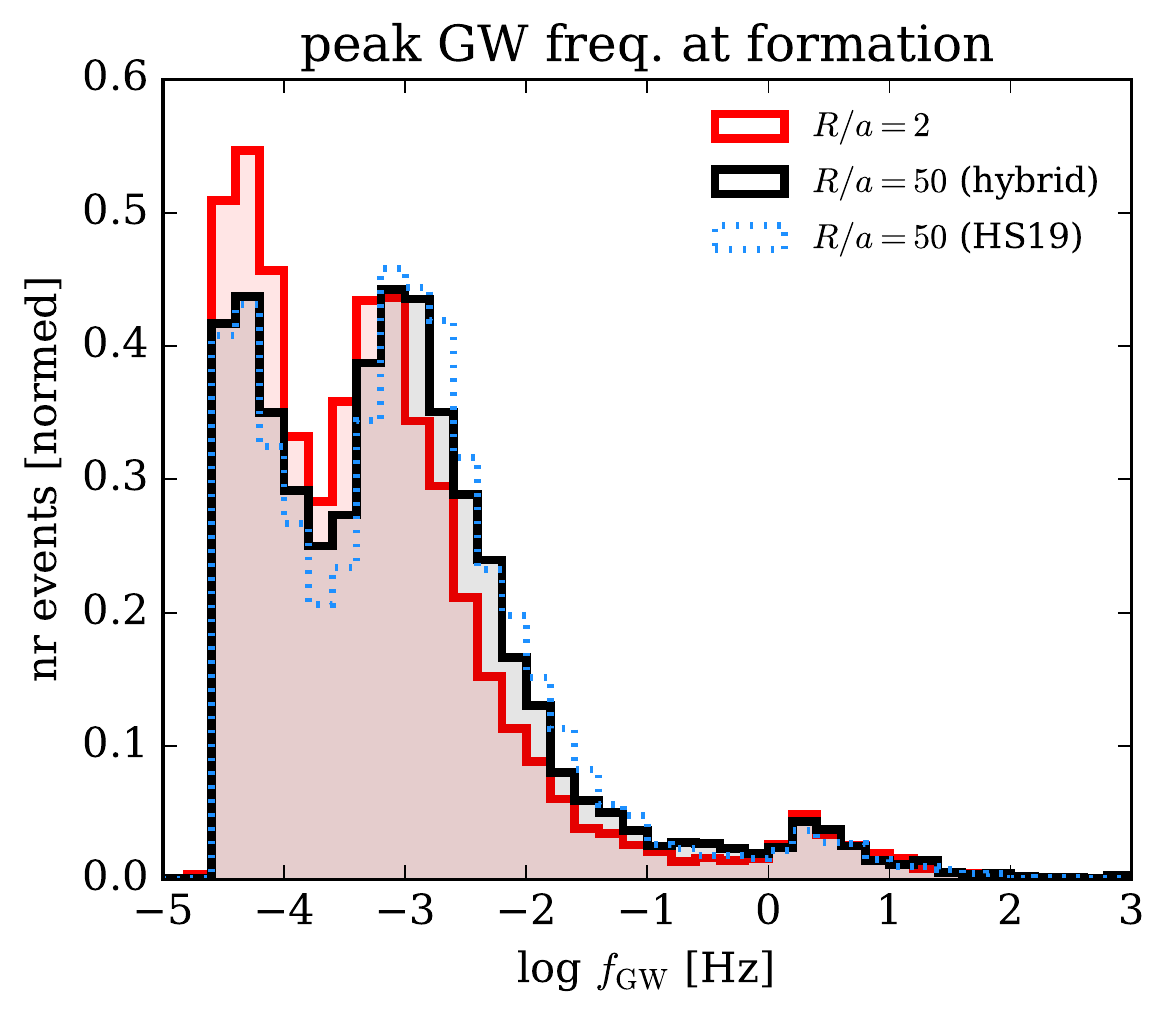}
\includegraphics[width=\columnwidth]{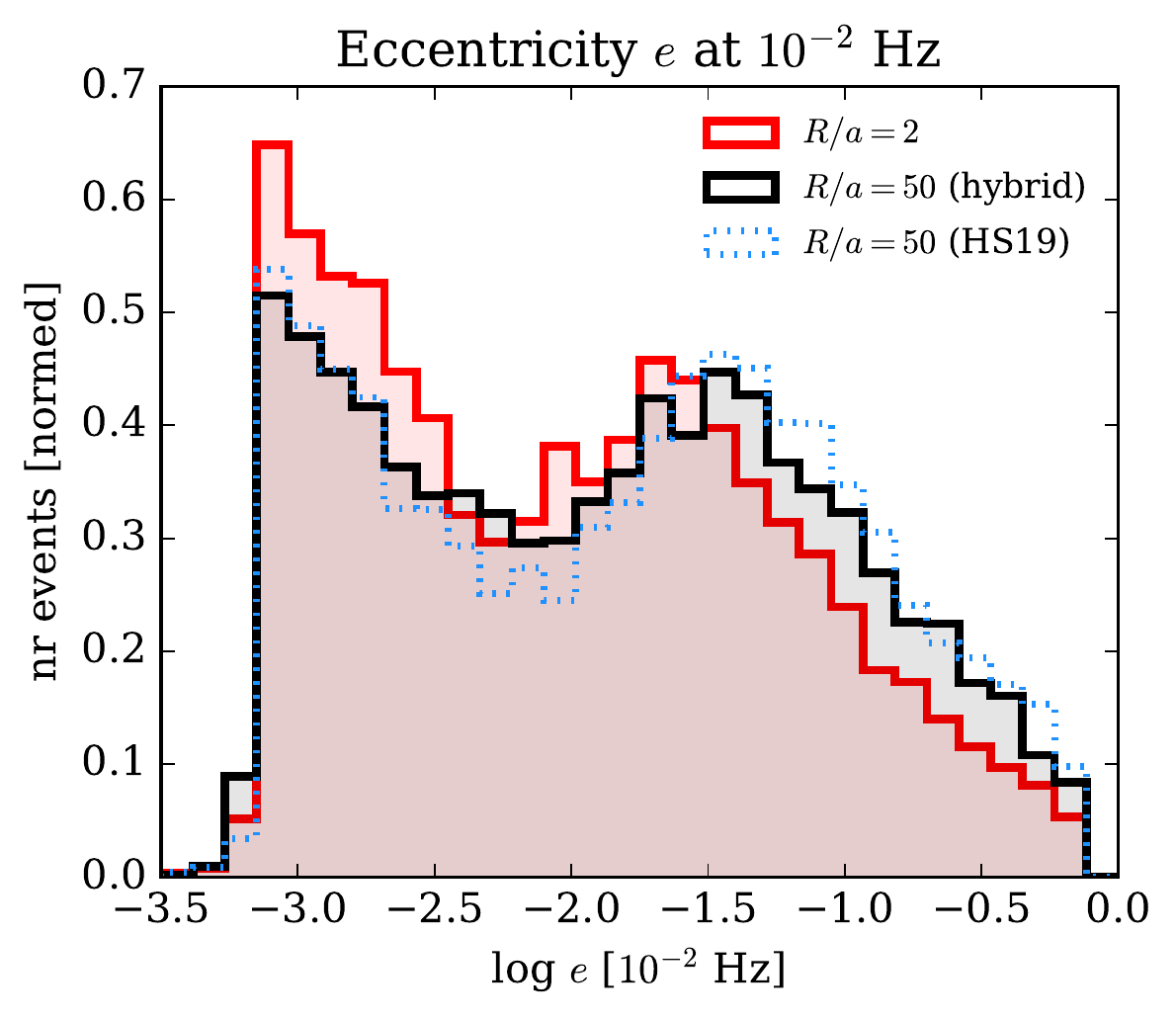}
\caption{Distribution of GW peak frequency $f_{\rm GW}$ ({\it top plot}), and
BBH orbital eccentricity evaluated at $f_{\rm GW} = 10^{-2}$ Hz ({\it bottom plot}), derived
from $15000$ individual runs of BBHs undergoing weak and strong scatterings
inside a dense cluster. For this, we used the numerical scheme outlined in
Section \ref{sec:BBH Evolution with Weak Interactions} and our `fiducial cluster values'
listed in Section \ref{sec:Results from Including Weak Encounters}.
The {\it solid red}, {\it solid black} and {\it dotted blue} histograms represent the following distributions, respectively.
{\it $R_{\rm p}/a = 2$}: Maximum peri-center distance is here set to $R_{\rm p}/a = 2$, i.e. this set includes
strong encounters only.
{\it $R_{\rm p}/a = 50$ (hybrid)}: Maximum peri-center distance is here set to $R_{\rm p}/a = 50$,
i.e. this set also includes weak encounters. To evolve and model each weak encounter we make use of our
`hybrid approach' described in Section \ref{sec:Results from Including Weak Encounters}.
{\it $R_{\rm p}/a = 50$ (HS19)}: Similar to $R_{\rm p}/a = 50$ (hybrid), but here all weak interactions are modeled
using HS19, i.e. the whole evolution is done using fully analytical prescriptions.
As seen, the full analytical
approach ($R_{\rm p}/a = 50$ (HS19)) provides are very good estimate compared to the more
accurate, but also much slower, hybrid approach ($R_{\rm p}/a = 50$ (hybrid)).
Associated cumulative distributions are shown in Fig. \ref{fig:CDISTe102}.
Results are discussed in Section \ref{sec:A Population of Black Hole Binaries}.
}
\label{fig:fGWe102SEWSE}
\end{figure}

\begin{figure}
\centering
\includegraphics[width=\columnwidth]{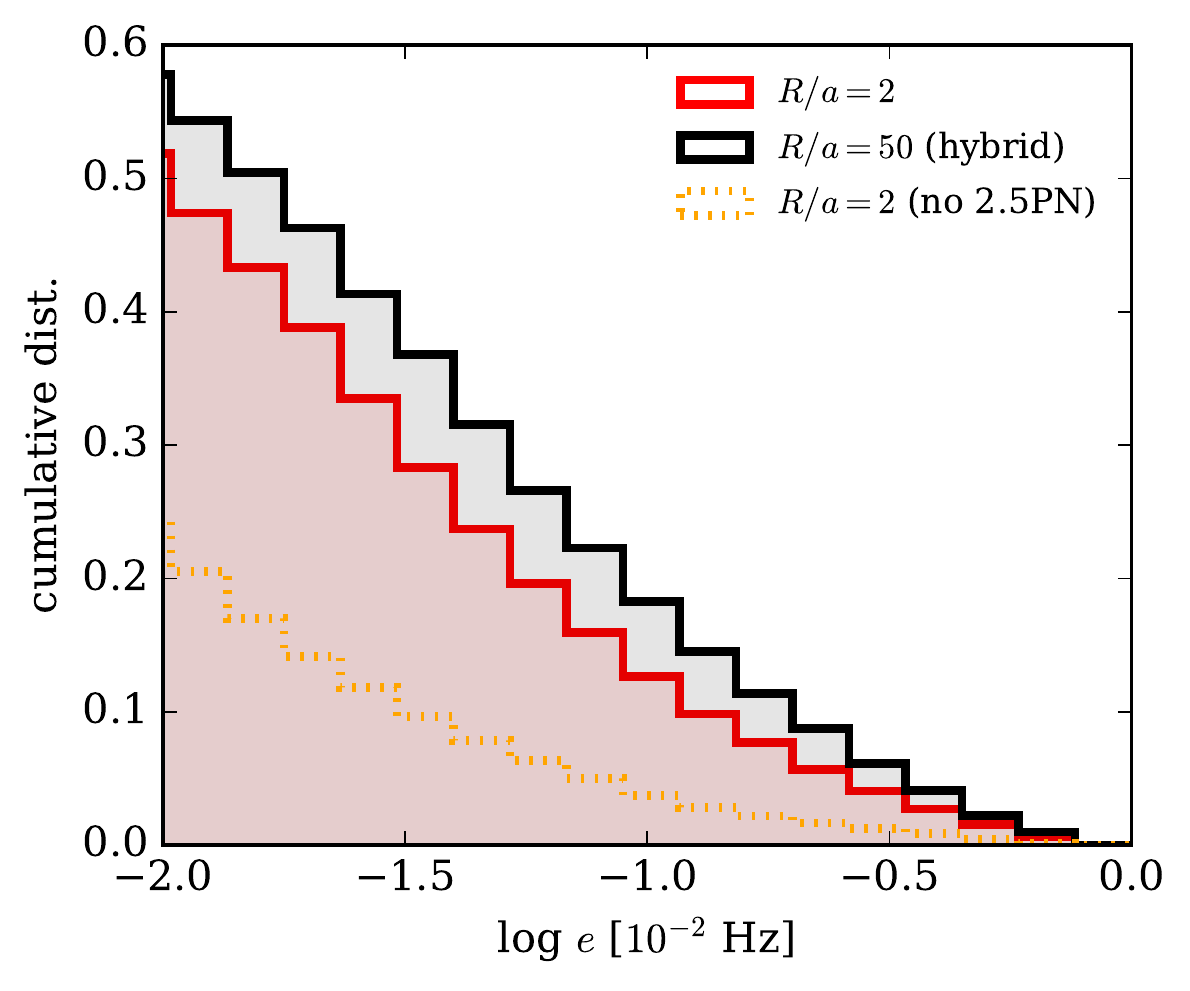}
\caption{Cumulative distribution of the bottom plot from Fig. \ref{fig:fGWe102SEWSE}.
The {\it solid red} and {\it solid black} distributions are described
in the caption of Fig. \ref{fig:fGWe102SEWSE}. The {\it dotted orange} line shows the result from
including strong encounters only and without any 2.5PN corrections, i.e.
it shows results from a pure Newtonian strong scattering approach.
As seen, for an accurate modeling one needs to include both 2.5PN corrects and weak encounters.
In \citep{2018MNRAS.tmp.2223S, 2018MNRAS.481.4775D} we pointed out for the first time the importance of 2.5PN corrections,
here we illustrate the importance of weak encounters.
}
\label{fig:CDISTe102}
\end{figure}

\begin{figure}
\centering
\includegraphics[width=\columnwidth]{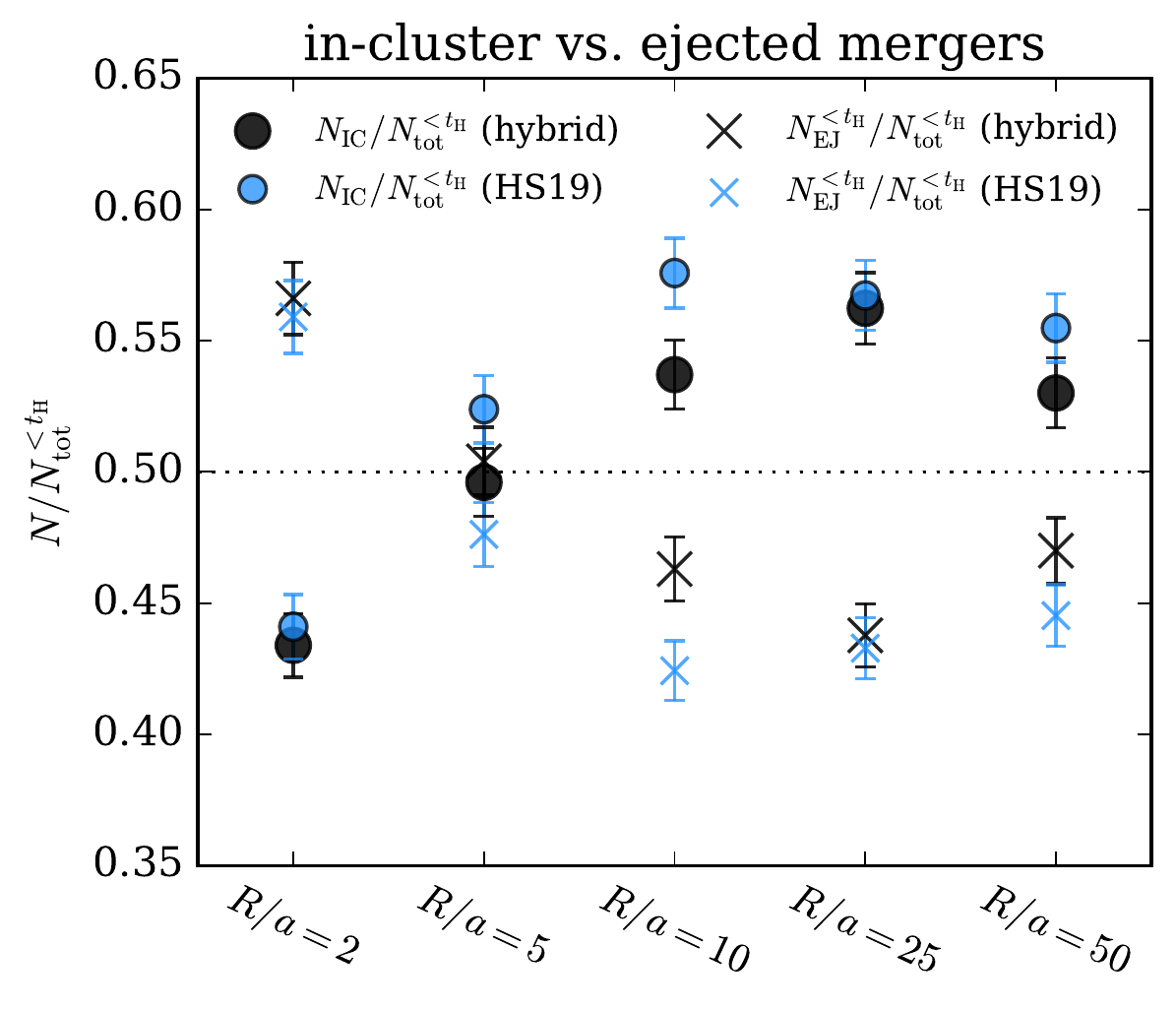}
\caption{Number of in-cluster BBH mergers ({\it circle}), $N_{\rm IC}$, and ejected BBHs with $\tau < t_{\rm H}$ ({\it cross}), $N_{\rm EJ}^{<t_{\rm H}}$,
each scaled by the total number of mergers, $N_{\rm tot}^{<t_{\rm H}} = N_{\rm IC} + N_{\rm EJ}^{<t_{\rm H}}$. Results are derived using our numerical scheme
outlined in Section \ref{sec:BBH Evolution with Weak Interactions} and our `fiducial cluster parameters' together
with our `hybrid approach' described in Section \ref{sec:Results from Including Weak Encounters}, for different values of $R_{\rm p}/a$ (x-axis).
The {\it black symbols} show results from our hybrid approach, where the {\it blue symbols} show results from our analytical MC approach.
As seen, if only strong encounters are included ($R_{\rm p}/a = 2$) the fractions $N_{\rm IC}/N_{\rm tot}^{<t_{\rm H}}$
and $N_{\rm EJ}^{<t_{\rm H}}/N_{\rm tot}^{<t_{\rm H}}$ are about $0.45$ and $0.55$, respectively, i.e. ejected BBH mergers dominate the merging population.
However, when we increase $R_{\rm p}/a$, and thereby start to include weak encounters, more and more of the BBHs tend to merge inside the cluster
before ejection becomes possible. Despite some Poisson scatter, this statement is consistent for both the hybrid and the analytical approach.
That the inclusion of weak interactions seems to increase the relative number of in-cluster mergers has major implications for, e.g.,
predicting accurate GW peak frequency and eccentricity distributions observable by especially LISA. It can also impact the number of
2. generation GW mergers.
}
\label{fig:NmergRp}
\end{figure}

To systematically explore the role of including weak encounters we now consider Fig. \ref{fig:fGWe102SEWSE}, which shows the
distribution of $f_{\rm GW}$ (top plot) and the eccentricity $e$ distribution at $10^{-2}$ Hz (bottom plot), where the {\it solid black} and {\it solid red} lines
show results from including weak interactions ($R_{\rm p}/a = 50$) and not including weak interactions  ($R_{\rm p}/a = 2$) in the BBH evolution, respectively.
The companion Fig. \ref{fig:CDISTe102} shows the cumulative distribution of the eccentricity distribution at $10^{-2}$ Hz from Fig. \ref{fig:fGWe102SEWSE}.
From considering these figures, one sees that including weak interactions leads to a notable increase in the number of GW sources that will
appear with measurable eccentricity ($>10^{-2}$) in the LISA band. For example, the number of sources with $\log(e) > -1.0$ at $10^{-2}$ Hz
increases by about $\approx 50\%$. The reason can be found from considering Fig. \ref{fig:NmergRp}, which shows the number
of in-cluster mergers (2-body + 3-body mergers) relative to the number of
BBHs ejected from the cluster with $\tau < t_{\rm H}$, where $t_{\rm H}$ denotes the Hubble time,
as a function of $R_{\rm p}/a$. As seen, as more and more weak encounters are included, i.e. as we
increase $R_{\rm p}/a$, more of the BBHs will merge inside the cluster before being ejected. The increased number of eccentric LISA sources
therefore primarily originates from an increase in the number of in-cluster mergers.
The corresponding slight systematic shift towards higher $e$ seen in the bottom plot of Fig. \ref{fig:fGWe102SEWSE} for the in-cluster mergers
also help increasing the number.

We do not yet have a solid analytical understanding for how the relative importance of weak interactions change with
the cluster parameters as we do for strong interactions \citep[e.g.][]{2018PhRvD..97j3014S}; however, the first results presented here are interesting, and weak interactions
definitely add an interesting complexity to the problem.
They are also highly important to include for an accurate modeling of
possible secular effects on the BBHs if the cluster is not perfectly spherical \citep{2019arXiv190201344H, 2019arXiv190201345H},
and might also play a key role in accurately predicting the formation rate of 2. generation mergers \citep[e.g.][]{2016ApJ...824L..12O}.
We reserve several of these questions for future work.

\subsection{Fast Analytical Monte Carlo Approach}\label{sec:Full Analytical Approach}

We finish this section by exploring how well one can do by using Eq. \eqref{eq:de2ord} for all weak interactions, i.e. for
all $r_{\rm p}/a_0 > \mathscr{C} (=2)$, instead of employing our hybrid approach that includes numerical simulations;
an approach we refer to as our `Analytical MC Approach'. In this approach it is unavoidable to
encounter situations where either $e>1$ or $e<0$, also when using the SO term, therefore, every time such a case appears we simply put $\de = 0$
without stopping the code. Results from this approach is shown in Fig.  \ref{fig:fGWe102SEWSE} and \ref{fig:NmergRp}.
As seen, despite some Poisson scatter, the Analytical MC Approach seems to perform very well. A great overlap between the
hybrid and the analytical MC approach is especially seen in Fig. \ref{fig:fGWe102SEWSE}, where both of them clearly distinguish themselves from
the runs where only strong interactions are included.

This is highly encouraging as the analytical MC approach is extremely fast and easy to code up,
which especially allows one to explore a wide range of cluster parameters in a very short amount of time. This approach is therefore also well suited for
performing a more systematic study on the importance of weak interactions as function of the properties of the cluster and the
interacting BHs.

Finally, there are several easy ways of improving our proposed analytical MC approach, two of which are:
$(1)$ For strong interactions use the full distribution of $\delta$ given e.g. in \citep{Heggie:1975uy}, instead of the average value $\delta = 7/9$.
$(2)$ For weak interactions, include the change in SMA $\Delta{a}$ using \citep{2003CeMDA..87..411R}, instead of simply assuming $\Delta{a} = 0$.
On top of these, it is possible to improve on the distribution of 3-body mergers, and also include 4-body mergers.
In an upcoming paper we will quantify the role of single-single GW capture mergers. All of these studies will become central
in developing the next generation of fast MC codes for modeling dense stellar systems and their formation of BBH GW mergers.

\section{Conclusions}\label{sec:Conclusions}

We have in this paper quantified the effects from weak encounters
on the assembly of BBH mergers in GCs using a combination of novel analytical
and numerical techniques. Current state-of-the-art codes \citep[e.g.][]{2018PhRvD..98l3005R, 2013MNRAS.431.2184G},
as well as recent analytical studies \citep[e.g.][]{2018MNRAS.tmp.2223S, 2018MNRAS.481.4775D}
have only included strong interactions in their modeling; however, BBHs in GCs will in-between each strong interaction
undergo hundreds of weak interactions, each of which changes the eccentricity
of the BBH slightly and thereby its GW inspiral life time. It is therefore possible for BBHs to both be driven towards
merger and also away from merger entirely through weak interactions.

Using a simple cluster model with the inclusion of weak and strong interactions,
we find here that the addition of weak interactions leads to a notable and systematic relative increase in the number
of BBHs merging in-side the cluster. For our fiducial cluster parameters
we find that the total number of observable BBH mergers goes from being dominated by ejected mergers to in-cluster mergers
when weak interactions are included. Because of the relative short GW inspiral time of in-cluster mergers, this
population will appear as eccentric LISA sources, and therefore add an important piece to the puzzle of
how to disentangle BBH merger channels from each other by the use of GW observations.

Our main results are based on a hybrid approach where we combine full 2.5PN numerical
scattering experiments with analytical expressions. We have in this regard
especially demonstrated that very similar results are achieved if one uses our new SO
perturbative solution \citep{2019arXiv190409624H} for modeling the changes in BBH
eccentricity for all weak interactions.
This is highly encouraging, as this allows for a full analytical MC modeling of the problem
including both weak and strong interactions as well as 2.5PN effects. This makes it e.g. possible
to derive distributions of BBH merger properties in just a few seconds for any cluster parameters. Of course,
this approach is based on a simplified cluster model, but previous work based on the same
assumptions turns out to be in surprisingly good agreement with more sophisticated MC models \citep[e.g.][]{2019arXiv190102889S}.

We have in this paper only considered results from a single set of cluster parameters, but we plan in the near
future to explore the relative importance of weak interactions for a range of cluster parameters
from less dense to very dense nuclear star clusters. It might be that in some regimes weak interactions
play an even larger role than found in this paper; however, we reserve that for future work.

Finally, we note that the weak interactions we have here considered are essential to include
for studying the secular evolution of BBHs in e.g. non-spherical clusters \citep{2019arXiv190201344H, 2019arXiv190201345H}.
The reason is that weak interactions not only give rise to a change in eccentricity, but also to a change in the orbital
orientation. This change can interfere with the secular change induced from a non-spherical potential,
and thereby affect the distribution of BBHs driven to merger in-side clusters through this mechanism.
Whether or not this is a competing channel for forming eccentric LISA sources will be studied in upcoming work.

{\it Acknowledgments. ---}
We thank Stephen McMillan and Piet Hut for stimulating discussions.
J.S. acknowledges support from the Lyman Spitzer Fellowship.
A.S.H. gratefully acknowledges support from the Institute for Advanced Study, and the Martin A. and Helen
Chooljian Membership.

\bibliographystyle{h-physrev}
\bibliography{NbodyTides_papers}

\end{document}